\documentclass[aps,prl,
twocolumn,
nofootinbib,
citeautoscript,
superscriptaddress,showkeys,showpacs]{revtex4-1} 
\usepackage{amsmath,amsfonts,amssymb}
\usepackage{color}
\usepackage[dvipsnames,table]{xcolor}
\usepackage{graphicx}
\usepackage[colorlinks=true, allcolors = magenta]{hyperref}
\graphicspath{{graphics/}}

\renewcommand{\vec}{\boldsymbol}

\newcommand{\m}[1]{\begin{pmatrix} #1 \end{pmatrix}}

\DeclareMathOperator{\tr}{tr}
\definecolor{mygold}{HTML}{edc812}
\definecolor{myred}{HTML}{d60021}
\definecolor{myblue}{HTML}{003bc6}
\renewcommand{\section}[1]{\noindent \emph{#1:}}

\begin{document}

\title{Symmetry-mixed bound-state order: extended degeneracy of $(d + ig)$-superconductivity in $\mathrm{Sr}_{2}\mathrm{Ru}\mathrm{O}_{4}$}

\author{Roland Willa}
\affiliation{Institute for Theory of Condensed Matter, Karlsruhe Institute of
Technology, Karlsruhe, Germany}

\begin{abstract}
We report a fluctuation-driven state of matter that develops near an accidental degeneracy point of two symmetry-distinct primary phases. Due to symmetry mixing, this bound-state order exhibits unique signatures, incompatible with either parent phase. 
Within a field-theoretical formalism, we derive the generic phase diagram for system with bound-state order, study its response to strain, and evaluate analytic expressions for a specific model.
Our results support the $(d + ig)$-superconducting state as a candidate for $\mathrm{Sr}_{2}\mathrm{Ru}\mathrm{O}_{4}$: Most noticeably, the derived strain-dependence is in excellent agreement with recent experiments [Hicks \emph{et al.} Science (2014) and Grinenko \emph{et al.} arXiv (2020)]. The evolution above a non-vanishing strain from a joint onset of superconductivity and time-reversal symmetry-breaking to two split phase transitions provides a testable prediction for this scenario.
\end{abstract}
\maketitle

%\section{Introduction}
The phase space of correlated matter is rich as electronic, magnetic, and structural degrees compete for dominance  \cite{deMedici2009, Korshunov2014, Boehmer2014, Boehmer2015, Maiti2015, Moll2015, Kim2018, Maier2019, Li2019, WillaK2019a, Agterberg2020}. Surprisingly often these orders are nearly degenerate such that a tuning parameter---like pressure, chemical doping, or magnetic field---allows to change the energy balance in favor of a different ground-state. This possibility is very pronounced in multiband systems such as the iron-based materials \cite{Graser2009, Fernandes2014, Wang2019, Fernandes2019}, heavy-fermion systems \cite{Okazaki2011, Chandra2013, Kung2015}, and different oxide families \cite{Emery1995, Chakravarty2001, Raghu2010, Scaffidi2014, Steppke2017, Zhao2015, Fechner2016, Pustogow2019, Ramires2019, Roising2019, Mackenzie2020}.
It is not immediately clear how to systematically treat the collision of electronic phases as their mutual interaction can be repulsive, attractive, or largely indifferent. Also, symmetry arguments for the classification of low-temperature phases are often obstructed by the possibility of nearly-degenerate phases. This may be particularly relevant for the decades-old quest of understanding superconductivity in $\mathrm{Sr}_{2}\mathrm{Ru}\mathrm{O}_{4}$, which has transitioned from an odd-parity $p$-wave superconductor \cite{Luke1998} to the recently proposed time-reversal symmetry-breaking even-parity $(d + id)$- \cite{Pustogow2019} or $(d + ig)$-superconductor \cite{Kivelson2020-condmat, Ghosh2020-condmat} with nearly degenerate $d_{xz}$/$d_{yz}$ and $d_{x^{2}-y^{2}}$/$g_{xy(x^{2}-y^{2})}$ states, respectively. As the last candidate gains further experimental support \cite{Ghosh2020-condmat}, it remains to be understood why two distinct order parameters coincidentally appear at a degeneracy point.

\begin{figure}[t]
\centering
\includegraphics[width = 0.48\textwidth]{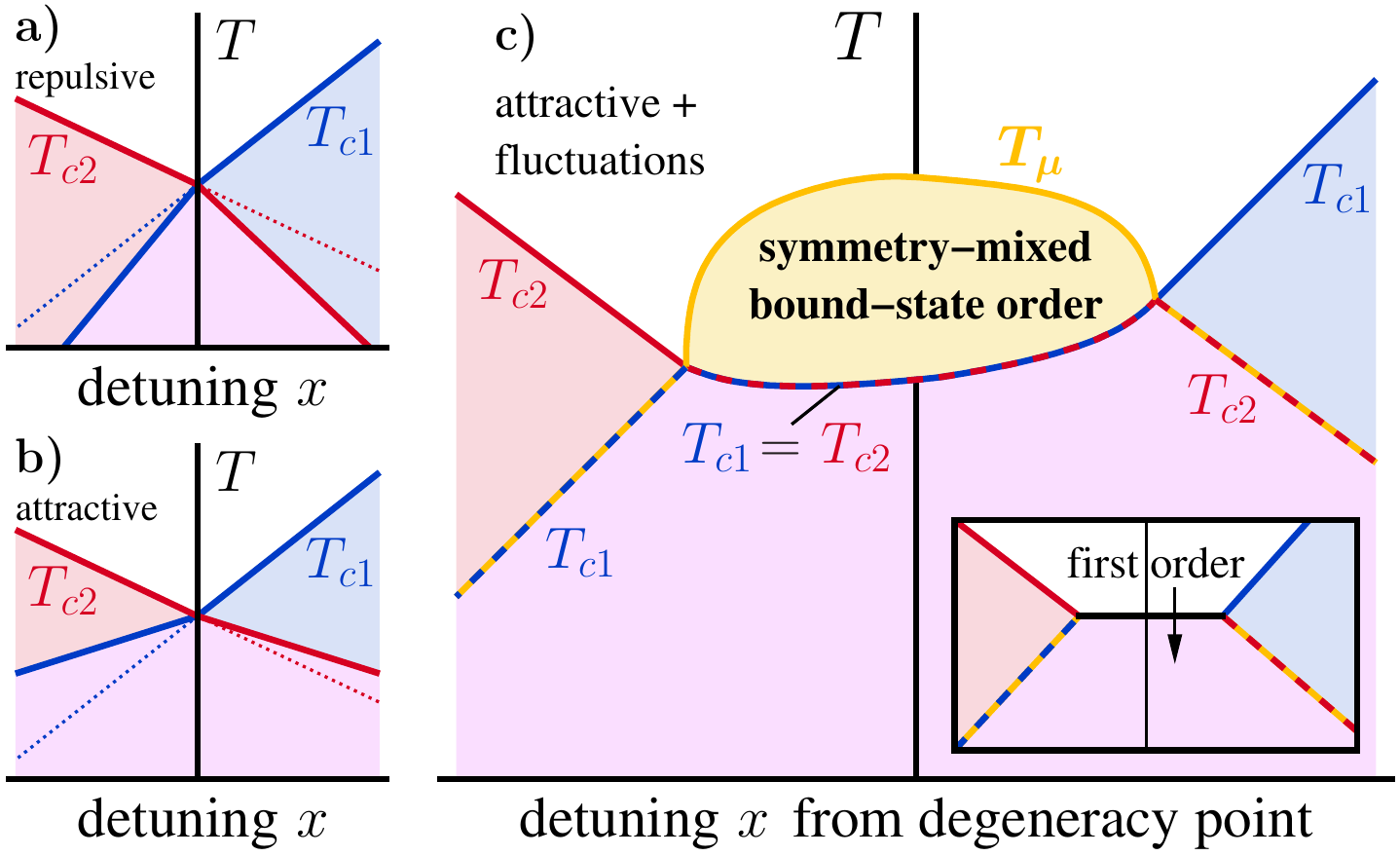}
\caption{Phase diagram of two phases $\langle \eta_{1} \rangle$ (blue) and $\langle \eta_{2} \rangle$ (red) near their degeneracy point $x \!=\! 0$ for repulsive [a)] or attractive [b)] phase interaction. Dotted lines indicate the phase boundary of the subordinate order in the absence of interaction. Including fluctuations [c)] the symmetry-mixed bound-state order $\mu = \langle -i (\eta_{1}^{*} \eta_{2} - \eta_{1} \eta_{2}^{*})/2 \rangle$ emerges in the vicinity of the phase intersection (yellow). This stand-alone phase acquires a finite expectation \emph{without} the appearance of the phases $\langle \eta_{j} \rangle$.
For sufficiently strong phase attraction the bound-state order gives way for a first-order transition into a fully ordered state (inset).
\label{fig:sketch}
}
\end{figure}

A standard and seemingly reasonable approach to tackle correlated phases of matter consists in refining the analysis to symmetry sectors defined by the known parent order parameters. The latter transform according to their irreducible representation and dictate which symmetries are broken at a phase transition.
This Letter introduces a fluctuation-driven phase of matter where two \emph{symmetry-distinct} phases $\eta_{1}$ and $\eta_{2}$ form a two-order bound-state, see Fig.\ \ref{fig:sketch}, while both primary phases remain absent. For real fields $\eta_{1}$ and $\eta_{2}$ the bound-state order parameter is $\mu \!=\! \langle \eta_{1} \eta_{2} \rangle$. In the case of two complex phases, it depends on the system's fluctuation-free ground state. If the latter breaks time-reversal symmetry [($\eta_{1} + i \eta_{2}$)], the bound-state $\mu \!=\! \langle -i (\eta_{1}^{*} \eta_{2} - \eta_{1} \eta_{2}^{*})/2 \rangle$ is also odd under time reversal, whereas it assumes the form $\mu \!=\! \langle (\eta_{1}^{*} \eta_{2} + \eta_{1} \eta_{2}^{*})/2\rangle$ in the even case. With this new order parameter transforming according to its own irreducible representation, the associated bound-state breaks a set of symmetries that is distinct from each parent phase. As described below, a bound-state order naturally appears near the accidental degeneracy of two complex phases.
This observation provides crucial support towards the $(d + ig)$-scenario \cite{Kivelson2020-condmat} for $\mathrm{Sr}_{2}\mathrm{Ru}\mathrm{O}_{4}$: While the appearance of a stand-alone bound-state phase may be difficult to detect, it's existence forces the two primary phases to appear jointly, thereby extending the accidental degeneracy point to a (near-)degeneracy line. Furthermore, the bound-state order qualitatively accounts for the strain dependence reported in Refs.\ \cite{Hicks2014, Watson2018, Grinenko2020-condmat}.

The formation of a bound-state order in low-dimensional systems is reminiscent of the vestigial orders discussed in the iron pnictide materials, which have successfully captured the nematic phases in proximity of spin-density wave instabilities \cite{Fang2008, Xu2008, Fernandes2010, Cano2010, Fernandes2012a, Fernandes2012b, Stanev2013, Willa2019a}. This concept has also been considered for symmetry-related $p_{x} \!\pm\! ip_{y}$ superconductors \cite{Fischer2016}. In that context, the two order parameters belong to the same irreducible representation which imposes additional constraints on the phenomenology. Here, in contrast, the primary orders generically develop at different transition temperatures and allow to study a possible formation of a bilinear bound-state away from the phase degeneracy.

Within a Ginzburg-Landau description, the free energy density of two nearly-degenerate complex order parameters $\eta_{1}$ and $\eta_{2}$ can be cast in terms of an expansion including all symmetry-allowed contributions.
Keeping the qualitative discussion on a general level, specific physical implications will be discussed for the superconducting orders $\eta_{1} \sim d_{x^{2}-y^{2}}$ and $\eta_{2} \sim g_{xy(x^{2}-y^{2})}$ proposed for $\mathrm{Sr}_{2}\mathrm{Ru}\mathrm{O}_{4}$ \cite{Kivelson2020-condmat}.
The problem's free energy density takes the form
\begin{align}
   \mathcal{F} &= 
      \frac{r_{0}}{2} (\eta^{*}_{1} \eta^{\phantom{*}}_{1} \!+\! \eta^{*}_{2} \eta^{\phantom{*}}_{2})
      \!-\! \frac{x}{2} (\eta^{*}_{1} \eta^{\phantom{*}}_{1} \!-\! \eta^{*}_{2} \eta^{\phantom{*}}_{2})
      \!+\! \frac{u_{+}}{8} (\eta^{*}_{1} \eta^{\phantom{*}}_{1} \!+\! \eta^{*}_{2} \eta^{\phantom{*}}_{2})^{2}
      \nonumber \\
      \label{eq:free-energy-density}
      &
      \quad
      + \frac{u_{-}}{8} (\eta^{*}_{1} \eta^{\phantom{*}}_{1} \!-\! \eta^{*}_{2} \eta^{\phantom{*}}_{2})^2
      - \frac{g}{8} [-i(\eta^{*}_{1} \eta^{\phantom{*}}_{2} \!-\! \eta^{\phantom{*}}_{1} \eta^{*}_{2})]^2
      + \mathcal{F}_{\nabla}\!\!
\end{align}
with $u_{\pm} = u \pm (g+\lambda)$, the phenomenological interaction parameters $u$, $g$, and $\lambda$, and a quadratic form $\mathcal{F}_{\nabla}$ of gauge-invariant gradient terms. A discussion of the possible ground-states is provided in Ref.\ \cite{Kivelson2020-condmat}. Focusing on the case of interest, $u_{\pm}, g \!>\! 0$ provides a ground-state that breaks time-reversal symmetry, where $\eta_{1}$ and $\eta_{2}$ have a relative phase shift of $\pm\pi/2$. The two phases are degenerate for $x \!=\! 0$, hence $x$ provides a natural tuning parameter away from that point. As, by assumption, the two order parameters belong to different irreducible representations, terms $\propto (\eta^{*}_{1} \eta^{\phantom{*}}_{2} \!\pm\! \eta^{\phantom{*}}_{1} \eta^{*}_{2})$ are symmetry-forbidden.

The assumptions $u \!\equiv\! u_{\pm}$ and $\mathcal{F}_{\nabla} \!=\! \frac{1}{2} \sum_{j}(\nabla \eta^{*}_{j})(\nabla \eta^{\phantom{*}}_{j})$ significantly simplify the results, without affecting the qualitative findings. When appropriate, non-trivial effects of relaxing these constraints shall be mentioned.

The spontaneous condensation of the bound-state order parameter necessitates an attractive interaction channel between the parent phases (here provided by $g \!>\! 0$).
%
%This contrasts to real fields that commonly repel and require a fluctuating degrees of freedom (e.g. phonons) to induce an effective attraction.
%
For the candidate $(d + ig)$-superconductor, the parent phase $d_{x^{2}-y^{2}}$ [$g^{\phantom{*}}_{xy(x^{2}-y^{2})}$] belongs to the $B_{1g}$ ($A_{2g}$) representation of the $D_{4h}$ point group, see Table~\ref{tab:product-table-D4h}. In this case the bound-state $\mu = \langle -i(\eta^{*}_{1} \eta^{\phantom{*}}_{2} \!-\! \eta^{\phantom{*}}_{1} \eta^{*}_{2})/2 \rangle$ belongs to the $B_{2g}$ representation. It preserves the $U(1)$ gauge
and is odd under time reversal.

\begin{table}
\setlength{\tabcolsep}{2.9pt}
\newcommand{\colcellone}[1]{\multicolumn{1}{| >{\columncolor{myblue!15}[\tabcolsep]} l|}{#1}}
\newcommand{\colcelltwo}[1]{\multicolumn{1}{| >{\columncolor{myred!15}[\tabcolsep]} l|}{#1}}
\newcommand{\colcellnem}[1]{\multicolumn{1}{| >{\columncolor{mygold!25}[\tabcolsep]} l|}{#1}}
\begin{tabular}{|l|c|c|c|c|c|c|c|c|c|c|}
\hline
$\Gamma \downarrow$         & $\,E\,$ & $2C_{4}^{z}$ & $C_{2}^{z}$ & $2C'_{2}$ & $2C''_{2}$ & $\,I\,$ &
$2 I C_{4}^{z}$ & $ I C_{2}^{z}$ & $2 I C'_{2}$ & $2 I C''_{2}$
\\ \hline
$\vec{A_{1g}}$ & 1 & 1 & 1 & 1 & 1 & 1 & 1 & 1 & 1 & 1 \\ \hline
\colcelltwo{$\vec{A_{2g}}$} & 1 & 1 & 1 & -1 & -1 & 1 & 1 & 1 & -1 & -1 \\ \hline
\colcellone{$\vec{B_{1g}}$} & 1 & -1 & 1 & 1 & -1 & 1 & -1 & 1 & 1 & -1 \\ \hline
\colcellnem{$\vec{B_{2g}}$} & 1 & -1 & 1 & -1 & 1 & 1 & -1 & 1 & -1 & 1 \\ \hline
\end{tabular}
\vspace{.5em}

\setlength{\tabcolsep}{4.9pt}
\renewcommand{\colcellone}[1]{\multicolumn{1}{ >{\columncolor{myblue!15}[\tabcolsep]} l|}{#1}}
\renewcommand{\colcelltwo}[1]{\multicolumn{1}{| >{\columncolor{myred!15}[\tabcolsep]} l|}{#1}}
\renewcommand{\colcellnem}[1]{\multicolumn{1}{ >{\columncolor{mygold!25}[\tabcolsep]} l|}{#1}}
\begin{tabular}{|l| c|c|c|c|}
\hline
$\Gamma(\eta_{2}) \downarrow$ \quad $\Gamma(\eta_{1}) \rightarrow$& $\vec{A_{1g}}$ & $\vec{A_{2g}}$ &  \colcellone{$\vec{B_{1g}}$} & $\vec{B_{2g}}$ \\ \hline
\qquad $\vec{A_{1g}}$ & $A_{1g}$ & $A_{2g}$ & $B_{1g}$ & $B_{2g}$\\ \hline
\colcelltwo{\qquad $\vec{A_{2g}}$} & $A_{2g}$ & $A_{1g}$ & \colcellnem{$B_{2g}$} & $B_{1g}$\\ \hline
\qquad $\vec{B_{1g}}$ & $B_{1g}$ & $B_{2g}$ & $A_{1g}$ & $A_{2g}$\\ \hline
\qquad $\vec{B_{2g}}$ & $B_{2g}$ & $B_{1g}$ & $A_{2g}$ & $A_{1g}$\\ \hline
\end{tabular}
\caption{
Excerpt of the character table (top) and the product table (bottom) for the $D_{4h}$ point group. The product table specifies to which irreducible representation $\Gamma(\eta_{1} \eta_{2})$ the bound-state of two primary phases $\eta_{j}$ belongs. The colored cells correspond to the relevant orders for $(d+ig)$ superconductivity in  $\mathrm{Sr}_{2}\mathrm{Ru}\mathrm{O}_{4}$ \cite{Kivelson2020-condmat}.
\label{tab:product-table-D4h}}
\end{table}
%

%\section{Generic treatment}
An established route to evaluate the condensation condition of a bound-state consists in replacing $\eta_{j}$ by an $N$-dimensional vector field $\vec{\eta}_{j}$ and treating the above action in a large-$N$ limit. With the identity,
\begin{align}
   1 = \!\int\! \mathcal{D}\sigma_{m} \mathcal{D}\phi_{m} \exp\Big\{ \!-\!\!\int_{\vec{q}}\! i \sigma_{m}(\vec{q}) [\phi_{m}(\vec{q}) - m(\vec{q})] \Big\}
\end{align}
interaction terms are brought to a quadratic form in the primary fields, where $\phi_{m}$ is a generalized Hubbard-Stratonovich field associated to a real field $m$. The functional integration over $\sigma_{m}$ imposes $\delta(\phi_{m} - m)$. The notation $\int_{\vec{q}}$ abbreviates the momentum-integral and accounts for the system's anisotropy. With the three Hubbard-Stratonovich fields $\phi_{j} = \eta_{j}^{2}$ and $\mu = -i(\eta^{*}_{1} \eta^{\phantom{*}}_{2} \!-\! \eta^{\phantom{*}}_{1} \eta^{*}_{2})/2$, the partition function transforms to
\begin{align}\label{}
   &Z \propto \int
         \mathcal{D}\sigma_{1}\mathcal{D}\sigma_{2}\mathcal{D}\sigma_{\mu}
         \mathcal{D}\phi_{1}\mathcal{D}\phi_{2}\mathcal{D}\mu
         \;
         e^{-N \mathcal{S}^{\mathrm{eff}}}, \quad \text{with}\\
   &\mathcal{S}^{\mathrm{eff}} =
      \!\sum_{i,j= \{1,2\}}\!\!\!\!  \langle\eta^{*}_{i}\rangle (\mathcal{G}_{0}^{-1})_{ij} \langle\eta^{\phantom{*}}_{j}\rangle
      +  \int_{\vec{q}} \tr[\log(\mathcal{G}_{\vec{q}}^{-1})]\\
      \nonumber
      &\hspace{2.7em} + \frac{u}{4} (\phi_{1}^{2} + \phi_{2}^{2}) - \frac{g}{2} \mu^{2} + i \sigma_{1} \phi_{1} + i \sigma_{2} \phi_{2} + 2 i \sigma_{\mu} \mu,
\end{align}
and
\begin{align}\label{}
   \mathcal{G}_{\vec{q}}^{-1} \!=\! \frac{1}{2}\m{
     r_{0} - x + \vec{q}^{2} - 2 i \sigma_{1}\!\! &  i (2 i \sigma_{\mu})\\
     -i (2 i \sigma_{\mu})   &  r_{0} + x + \vec{q}^{2} - 2i \sigma_{2} }.
\end{align}
Here, the quadratic fluctuations of the fields $\eta_{j}$ around the mean value $\langle \eta_{j} \rangle$ have been integrated out. Furthermore, $\phi_{j}$ and $\mu$ are assumed uniform in space. The above action captures all possibilities of forming primary and/or bound-state phases \cite{Hecker2020-thesis}.
The large-$N$ limit now allows to search for saddle-point solutions to the remaining fields. Minimizing the action for $\phi_{j}$ and $\mu$ provides $2 i \sigma_{j} \!=\! - u \phi_{j}$ and $2 i \sigma_{\mu} \!=\! g \mu$.
The saddle-point conditions then read
\begin{align}
   \label{eq:eq1-general}
   r_{1} &= r_{0,1} + u \langle \eta^{*}_{1}\rangle \langle \eta^{\phantom{*}}_{1}\rangle + 2u \!\int_{\vec{q}} \frac{ r_{2} + \vec{q}^{2} }{ (r_{1} +  \vec{q}^{2}) (r_{2} +  \vec{q}^{2}) - (g \mu)^{2}},\\
   \label{eq:eq2-general}
   r_{2} &= r_{0,2} + u \langle \eta^{*}_{2}\rangle \langle \eta^{\phantom{*}}_{2}\rangle + 2u \!\int_{\vec{q}} \frac{ r_{1} + \vec{q}^{2} }{ (r_{1} +  \vec{q}^{2}) (r_{2} +  \vec{q}^{2}) - (g \mu)^{2}},\\
   \label{eq:eqmu-general}
   2\mu &= i(\langle\eta^{\phantom{*}}_{1}\rangle \langle\eta^{*}_{2}\rangle \!-\! \langle\eta^{*}_{1}\rangle \langle\eta^{\phantom{*}}_{2}\rangle)
         + \!\int_{\vec{q}} \frac{ 4g \mu }{ (r_{1} \!+\!  \vec{q}^{2}) (r_{2} \!+\!  \vec{q}^{2}) \!-\! (g \mu)^{2}},
\end{align}
where $r_{0,1} \!=\! r_{0} \!-\! x$, $r_{0,2} \!=\! r_{0} \!+\! x$ are the bare, and $r_{j} \!\equiv\! r_{0,j} + 2 u \phi_{j}$ the fluctuation-renormalized masses. The possibility of having non-zero primary phases imposes the additional constraints \cite{Hecker2020-thesis}
\begin{align}
   \label{eq:constraint-1}
   r_{1} \langle \eta_{1} \rangle &= - i g \mu \langle\eta_{2}\rangle,\\
   \label{eq:constraint-2}
   r_{2} \langle \eta_{2 }\rangle &= i g \mu \langle\eta_{1}\rangle.
\end{align}

\begin{figure}[t]
\centering
\includegraphics[width = 0.48\textwidth]{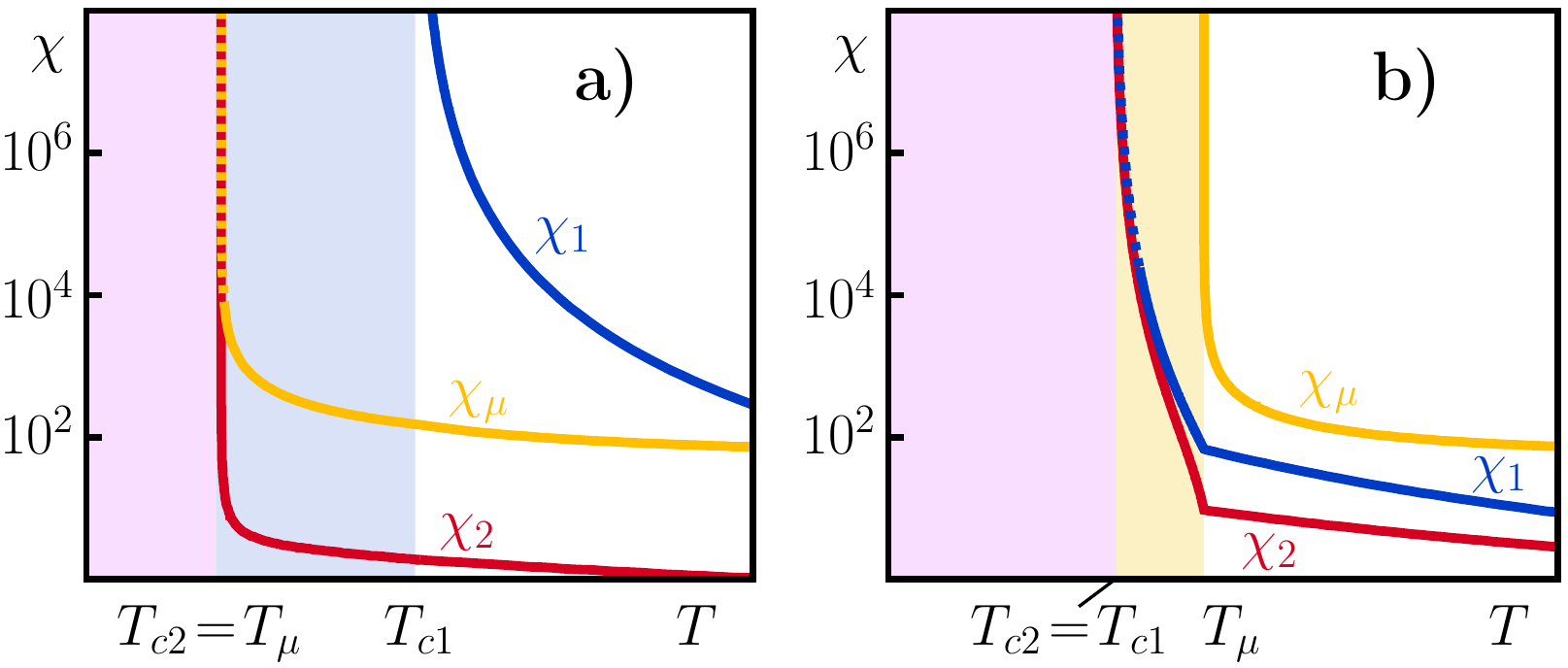}
\caption{
Susceptibilities $\chi_{1}(T)$ (blue), $\chi_{2}(T)$ (red), $\chi_{\mu}(T)$ (yellow) upon approaching the ordered phase. Far from the phase degeneracy [a)], the order $\langle \eta_{1} \rangle$ sets in first, followed by the instability in the two remaining sectors. Near the degeneracy [b)], the symmetry-mixed bound-state $\mu$ appears (yellow) while the parent orders remain zero. A renormalization of the susceptibilities induces a simultaneous divergence of $\chi_{1}$ and $\chi_{2}$. 
\label{fig:susceptibility}
}
\end{figure}

The susceptibility of the order parameters is evaluated by coupling conjugate fields $h_{j}$ and $h_{\mu}$ to the primary and bound-state orders. A physical implementation of $h_{\mu}$ is discussed below. Following a similar derivation as before one finds in the absence of primary phases
\begin{align}\label{eq:chi-j}
   \chi_{j} &\equiv \frac{\partial \langle \eta_{j} \rangle}{\partial h_{j}} \Big|_{h_{j}\to 0}
      = \frac{1}{r_{j}}\frac{r_{1}r_{2}}{r_{1} r_{2} - g^{2}\mu^{2}},\\
   \label{eq:chi-mu}
   \chi_{\mu} &\equiv \frac{\partial \mu}{\partial h_{\mu}} \Big|_{h_{\mu}\to 0}
   = \frac{\mathrm{K}_{\mu}(r_{1},r_{2}) }{1 - g \mathrm{K}_{\mu}(r_{1},r_{2})},
\end{align}
for the longitudinal susceptibilities, where
\begin{align}\label{}
   \mathrm{K}_{\mu}(r_{1},r_{2}) \equiv \int_{\vec{q}} \frac{2}{ (r_{1} +  \vec{q}^{2}) (r_{2} +  \vec{q}^{2})},
\end{align}
and $r_{1}$, $r_{2}$ satisfy Eqs.~\eqref{eq:eq1-general} and \eqref{eq:eq2-general} for $\mu \!=\! 0$. Once one primary phase develops [here $\langle \eta_{1} \rangle$] the remaining susceptibilities are modified to
\begin{align}\label{eq:chi-j-mod}
   \chi_{2} &= \frac{1}{r_{2}} \frac{
                                         1 - g \mathrm{K}_{\mu}(0,r_{2})}{
                                         1 - g \mathrm{K}_{\mu}(0,r_{2}) - g  \langle \eta^{*}_{1} \rangle \langle \eta^{\phantom{*}}_{1} \rangle / r_{2}}\\
   \label{eq:chi-mu-mod}
   \chi_{\mu} &= \frac{\mathrm{K}_{\mu}(0,r_{2}) + \langle \eta^{*}_{1} \rangle \langle \eta^{\phantom{*}}_{1} \rangle / r_{2}}{1 - g \mathrm{K}_{\mu}(0,r_{2}) - g \langle \eta^{*}_{1} \rangle \langle \eta^{\phantom{*}}_{1} \rangle / r_{2}}.
\end{align}
Contemplating Eqs.\ \eqref{eq:eq1-general}-\eqref{eq:constraint-2}---which determine the phase diagram for the three orders $\langle\eta_{1}\rangle$, $\langle\eta_{2}\rangle$, and $\mu$ and the two renormalized masses $r_{1}$ and $r_{2}$---allows to establish the following generic observations:

The bound-state order appears purely as a fluctuation phenomenon, as indicated by the momentum integrals coupling the different equations. As for the vestigial phases in iron-based systems, sufficiently strong fluctuations are required to trigger such a phase \cite{Fang2008, Xu2008, Fernandes2010, Cano2010, Fernandes2012a, Fernandes2012b, Stanev2013,  Willa2019a}. In fact, for large masses $r_{j}$ the integrals in Eqs.\ \eqref{eq:eq1-general}-\eqref{eq:eqmu-general} tend to be small. As a corollary, the bound-state phase is expected in the vicinity of a phase degeneracy where the renormalized masses $r_{1}$ and $r_{2}$ become small.

With each bilinear product of two of the three orders acting as a conjugate field to the third one, see Eqs.~\eqref{eq:eqmu-general}-\eqref{eq:constraint-2}, the presence of solely two orders is excluded. The phase diagram therefore exhibits at most two ordering transitions.
Starting from the high-temperature phase, the system can undergo a sequence of transitions into ordered phases following one of three scenarios:

\begin{figure}[t]
\centering
\includegraphics[width = 0.48\textwidth]{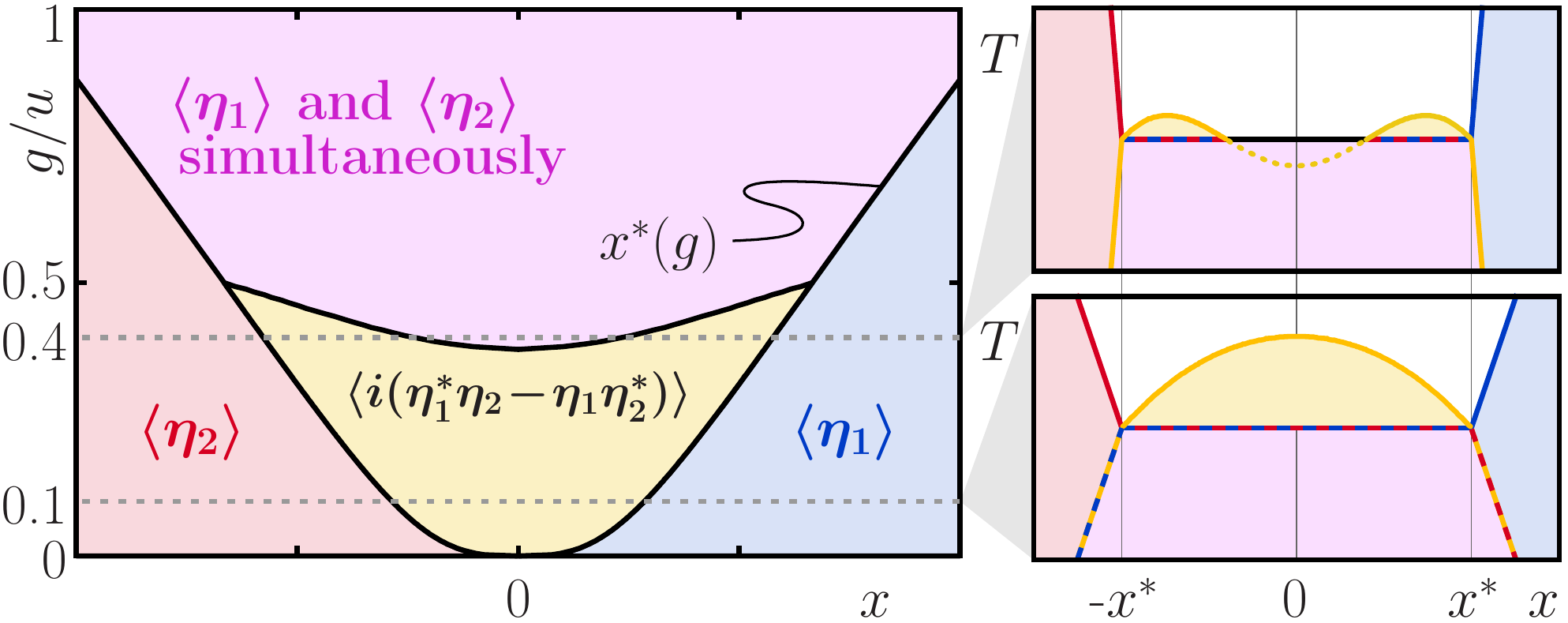}
\caption{
Left: In the phase space of two symmetry-distinct order parameters $\eta_{j}$ in the vicinity of the degeneracy point $x \!=\! 0$ and with an attractive interaction $g$, this color-map indicates which order appears first upon cooling: Far from the degeneracy $|x| > x^*(g)$, a primary order (either red or blue) appears through a second order transition. For large $g / u$ the system undergoes a first order transition (magenta) to a fully ordered state, where all orders become finite at once. For attractive interactions $g < u/2$ the symmetry-mixed bound-state $\mu = \langle -i(\eta^{*}_{1} \eta_{2} \!-\! \eta_{1} \eta^{*}_{2})/2 \rangle$ acquires a finite expectation value, \emph{without} the appearance of the primary phases (yellow). The right panels show the $Tx$-phase diagram for two specific interaction strengths $g/u = 0.4$ and $0.1$, respectively.
\label{fig:bso-phase-space}
}
\end{figure}

\medskip\noindent
\emph{Scenario 1: Ordering of a primary phase}
Far from a phase degeneracy one primary order appears with the subordinate one remaining zero. Let $\langle \eta_{1} \rangle$ be the phase to condensate first when $r_{1}$ vanishes. As phase interactions are still inactive, this transition line coincides with the conventional $T_{c1}$. If the above assumption $u_{+} =  u_{-}$ is relaxed, the transition is affected by a shift $\propto (u_{+}-u_{-})$. With fluctuations $T_{c1}$ is shifted down with respect to the bare $T_{c1}^{0}$ where $r_{0,1} \!=\! 0$. As the transition coincides with a divergent susceptibility $\chi_{1}$, Eq.\ \eqref{eq:chi-j}, it is of second order, see Fig.\ \ref{fig:susceptibility}a). At the phase boundary, the set of symmetries associated with irreducible representation $\Gamma(\eta_{1})$ is spontaneously broken.

Below $T_{c1}$, the primary phase $\langle \eta_{1} \rangle$ follows a typical behavior $\langle \eta_{1} \rangle \!\propto\! (T_{c1}-T)^{1/2}$, while $r_{1}$ is pinned to zero, see Eq.\ \eqref{eq:constraint-1}. The remaining orders appear simultaneously as the susceptibilities in Eqs.\ \eqref{eq:chi-j-mod} and \eqref{eq:chi-mu-mod} diverge. At this second-order transition further symmetries---according to the irreducible representation $\Gamma(\eta_{2})$---are broken.
The above ordering mechanism always precedes the onset driven by a disappearance of $r_{2}$ in Eq. \eqref{eq:chi-j-mod}.

\medskip\noindent
\emph{Scenario 2: Appearance of bound-state order}
By definition of the phase degeneracy, the two primary phases appear simultaneously at $x \!=\! 0$ in the absence of phase interactions. In the vicinity of that point both $r_{1}$ and $r_{2}$ are small and the bound-state $\mu$ can appear as a stand-alone phase. Its onset is determined as the bound-state susceptibility diverges, see Eq.\ \eqref{eq:chi-mu}, i.e.
\begin{align}\label{eq:mu-pure}
   1 - g \mathrm{K}_{\mu}(r_{1},r_{2}) = 0.
\end{align}
To allow for a non-zero solution to $\mu$ without the appearance of the primary phases, the balance between the right- and left-hand side of Eq.\ \eqref{eq:eqmu-general} must be guaranteed by a momentum integral. This highlights particularly well the fluctuation-driven origin of the bound-state phase. Upon entering this phase, the system breaks the symmetries associated with the irreducible representation $\Gamma(\eta_{1} \eta_{2})$. As a corollary and curious consequence, all point-group symmetries that are broken by \emph{both} primary orders remain preserved in the bound-state phase. For the superconducting $d_{x^{2}-y^{2}}$ and $g^{\phantom{*}}_{xy(x^{2}-y^{2})}$ states of $\mathrm{Sr}_{2}\mathrm{Ru}\mathrm{O}_{4}$, this observation applies to two-fold rotations about the diagonal axes [110] and [-110] (see $C''_{2}$ in Table~\ref{tab:product-table-D4h}). As the bound-state order breaks time-reversal symmetry it does not directly couple to the lattice and hence, is not a nematic state. At the same time it is also insensitive to a magnetic field. However, the order $\mu$ can be induced by applying an external magnetic field $H_{c}$ (along the crystallographic $c$ axis) \emph{together} with $B_{1g}$ strain through a coupling $\epsilon_{B_{1g}} H_{c} \mu$. This implies that $h_{\mu} = \epsilon_{B_{1g}} H_{c}$ is a conjugate field to $\mu$.

For a finite $\mu$, the primary phases now appear simultaneously when the criterion $r_{1} r_{2} = (g \mu)^{2}$ is first met, i.e., as both susceptibilities $\chi_{j}$ in Eq.\ \eqref{eq:chi-j} jointly diverge. The evolution of the three susceptibilities $\chi_{j}$ and $\chi_{\mu}$ is shown in Fig.\ \ref{fig:susceptibility}b), where the bound-state phase appears first and renormalizes the primary susceptibilities, Eq.\ \eqref{eq:chi-j}, to force a joint transition. At this second-order transition, both orders $\langle \eta_{j} \rangle$ develop and break the remaining symmetries associated with $\Gamma(\eta_{j})$. For the test case, this implies a full superconducting $(d + ig)$ state.
The bound-state is limited in phase space by a maximal attractive interaction strength $g^{*} = u/2$ and a maximal distance $x^{*}(g)$ away from the degeneracy point, see Fig.\ \ref{fig:bso-phase-space}. For larger interactions $g \!>\! g^{*}$, the line $x^{*}(g)$ separates between a joint transition (when $|x| \!<\! x^{*}$, scenario 3 below, or split transitions (when $|x| \!<\! x^{*}$, scenario 1 of the two primary phases. The Supplemental Material specifies the analytic model from which this phase space is computed.

\begin{figure}[t]
\centering
\includegraphics[width = 0.48\textwidth]{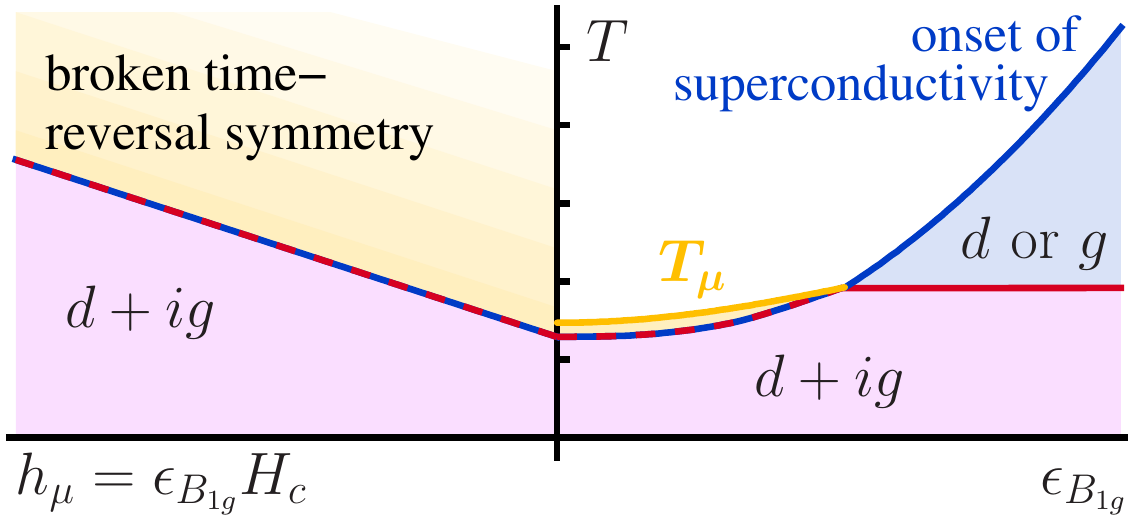}
\caption{Qualitative dependence of the transition temperatures for a $(d+ig)$-superconductor to an external $B_{1g}$-strain with (left) and without (right) magnetic field $H_{c}$. In the first case, the bound-state order is induced by $h_{\mu} = \epsilon_{B_{1g}} H_{c}$ and imposes a joint ordering of the $d$ and $g$ component. Without $H_{c}$, the system first develops a bound-state followed by a full $(d + ig)$-superconducting state. In this sequence the strain-dependence is indirect, i.e., through $r_{1}(\epsilon)$ in Eq.\ \eqref{eq:eqmu-general}. Above  critical strain the superconducting (blue) and the time-reversal-symmetry breaking (red) transitions split. The superconducting transition temperature now directly depends on $r_{1}(\epsilon)$ via Eq.\ \eqref{eq:eq1-general}.
\label{fig:strain}
}
\end{figure}

\medskip\noindent
\emph{Scenario 3: First order transition to fully ordered state}
As a third possibility, the system may undergo a first-order transition by discontinuously jumping to a finite $\mu$ and triggering a continuous appearance of $\langle \eta_{1} \rangle$ and $\langle \eta_{2} \rangle$. Here, the onset of all three phases is not accompanied by a divergent susceptibility. This scenario becomes relevant for large values $g/u$ in agreement with the simultaneous appearance of both primary phases obtained in a mean-field analysis, see also discussion below. In $\mathrm{Sr}_{2}\mathrm{Ru}\mathrm{O}_{4}$, the coherence lengths are relatively large and a quantitative distinction between scenario 2 (with a slim bound-state phase) may be experimentally challenging. Qualitatively however, both scenarios provide finite tuning range over which the breaking of the $U(1)$ gauge and time-reversal symmetry (almost) coincide, hence transforming the degeneracy point to an extended (near-)degeneracy line.

A meaningful discussion of the system's response to external strain requires specific assumptions on the phases $\langle \eta_{j} \rangle$. The following one is tailored to the test case $\mathrm{Sr}_{2}\mathrm{Ru}\mathrm{O}_{4}$. Noting that $A_{1g}$ strain merely rescales all couplings and $B_{2g}$ strain modifies the system's ground state away from a pure $(d+ig)$ case [by coupling to $(d^{*}g + d g^{*})$], the treatment is further narrowed to the interesting strain sector $B_{1g}$. While a bare linear coupling is prohibited by symmetry, strain couples to the bound-state order in combination with an external field $H_{c}$ along $c$: The combination $\epsilon_{B_{1g}} H_{c} \mu$ is symmetry-allowed and breaks time-reversal symmetry at any temperature. The number of phase transitions is then reduced to one, see Fig.\ \ref{fig:strain}, where both superconducting orders appear. In the absence of magnetic fields, $B_{1g}$ strain renormalizes the masses to $r_{j}(\epsilon_{B_{1g}}) \approx r_{j}(0) + r_{j}''(0) \epsilon_{B_{1g}}^{2}$. While it is possible to treat the generic case, it is instructive and reasonable to assume that strain dominantly couples to one primary order [we set $r_{1}''(0) < 0 $ and $r_{2}''(0) = 0$]. Near the degeneracy the system features two distinct regimes of strain response, see Fig.\ \ref{fig:strain}: At first, strain weakly affects the superconducting onset temperature [both with/without upstream bound-state order] through Eq. \eqref{eq:eqmu-general}. At larger strain, the system is pushed towards a split transition where the superconductivity precedes time-reversal symmetry breaking at a transition directly dictated by the strain-dependence of $r_{1}(\epsilon_{B_{1g}})$, see Eq.\ \eqref{eq:eq1-general}. This behavior is in excellent agreement with recent experiments \cite{Hicks2014, Watson2018, Grinenko2020-condmat}. As a falsifiable prediction, $\mu$SR experiments under $B_{1g}$ strain should resolve the splitting of an (almost) joint superconducting and time-reversal-symmetry breaking transition for strain below $\sim 0.05\%$ (based on the kink-like feature of $T_{c}$ observed in Ref.\ \cite{Hicks2014}) into two separate transitions for larger strain.

%\section{Conclusion}
In conclusion, this Letter presents a fluctuation-driven phase of matter that is the symmetry-mixed bound-state order $\langle -i (\eta_{1}^{*} \eta_{2} - \eta_{1} \eta_{2}^{*})/2 \rangle$ of two complex primary orders $\eta_{j}$. This order parameter naturally emerges as a stand-along phase near the degeneracy point of the parent orders. As this bound-state phase develops, it breaks a set of symmetries associated with the irreducible representation $\Gamma(\eta_{1}\eta_{2})$ rather than those dictated by the parent phases. Consequently the emergent phase preserves all point-group symmetries that are broken by both parent phases. The results presented here demonstrate that the symmetry sectors of parent phases give a too narrow view on the possible electronic states of matter in the vicinity of degeneracy points.

This phenomenology provides support for a $(d + ig)$-superconducting state in $\mathrm{Sr}_{2}\mathrm{Ru}\mathrm{O}_{4}$: While the exact degeneracy between the $d$- and $g$-component is rather improbable, fluctuations allow for a time-reversal symmetry breaking transition into a bound-state phase to precede (scenario 2) or coincide (scenario 3) with the superconducting transition over a finite tuning range away from the degeneracy point. Furthermore, the results under $B_{1g}$ strain exhibit two distinct regimes of parabolic strain dependence in good agreement with recent experiments \cite{Hicks2014, Watson2018, Grinenko2020-condmat}. As a testable consequence, the bound-state order should be observed in Kerr experiments \cite{Xia2006} when induced by simultaneously applying $B_{1g}$-strain and a magnetic field $H_{c}$ along $c$.

\begin{acknowledgments}
I gratefully thank R.\ Fernandes, M.\ Hecker, C.\ Hicks, S.A.\ Kivelson, J.\ Schmalian, C.\ Sp\r{a}nsl\"{a}tt, and K.\ Willa for stimulating discussions.
\end{acknowledgments}

\vfill
\onecolumngrid

%\bibliographystyle{apsrev4-1-titles}
%\bibliography{HiddenOrder}

\begin{thebibliography}{46}%
\makeatletter
\providecommand \@ifxundefined [1]{%
 \@ifx{#1\undefined}
}%
\providecommand \@ifnum [1]{%
 \ifnum #1\expandafter \@firstoftwo
 \else \expandafter \@secondoftwo
 \fi
}%
\providecommand \@ifx [1]{%
 \ifx #1\expandafter \@firstoftwo
 \else \expandafter \@secondoftwo
 \fi
}%
\providecommand \natexlab [1]{#1}%
\providecommand \enquote  [1]{``#1''}%
\providecommand \bibnamefont  [1]{#1}%
\providecommand \bibfnamefont [1]{#1}%
\providecommand \citenamefont [1]{#1}%
\providecommand \href@noop [0]{\@secondoftwo}%
\providecommand \href [0]{\begingroup \@sanitize@url \@href}%
\providecommand \@href[1]{\@@startlink{#1}\@@href}%
\providecommand \@@href[1]{\endgroup#1\@@endlink}%
\providecommand \@sanitize@url [0]{\catcode `\\12\catcode `\$12\catcode
  `\&12\catcode `\#12\catcode `\^12\catcode `\_12\catcode `\%12\relax}%
\providecommand \@@startlink[1]{}%
\providecommand \@@endlink[0]{}%
\providecommand \url  [0]{\begingroup\@sanitize@url \@url }%
\providecommand \@url [1]{\endgroup\@href {#1}{\urlprefix }}%
\providecommand \urlprefix  [0]{URL }%
\providecommand \Eprint [0]{\href }%
\providecommand \doibase [0]{http://doi.org/}%
\providecommand \selectlanguage [0]{\@gobble}%
\providecommand \bibinfo  [0]{\@secondoftwo}%
\providecommand \bibfield  [0]{\@secondoftwo}%
\providecommand \translation [1]{[#1]}%
\providecommand \BibitemOpen [0]{}%
\providecommand \bibitemStop [0]{}%
\providecommand \bibitemNoStop [0]{.\EOS\space}%
\providecommand \EOS [0]{\spacefactor3000\relax}%
\providecommand \BibitemShut  [1]{\csname bibitem#1\endcsname}%
\let\auto@bib@innerbib\@empty
%</preamble>
\bibitem [{\citenamefont {de' Medici}\ \emph {et~al.}(2009)\citenamefont {de'
  Medici}, \citenamefont {Hassan}, \citenamefont {Capone},\ and\ \citenamefont
  {Dai}}]{deMedici2009}%
  \BibitemOpen
  \bibfield  {author} {\bibinfo {author} {\bibfnamefont {L.}~\bibnamefont {de'
  Medici}}, \bibinfo {author} {\bibfnamefont {S.~R.}\ \bibnamefont {Hassan}},
  \bibinfo {author} {\bibfnamefont {M.}~\bibnamefont {Capone}}, \ and\ \bibinfo
  {author} {\bibfnamefont {X.}~\bibnamefont {Dai}},\ }\bibfield  {title} {\emph
  {\bibinfo {title} {{Orbital-Selective Mott Transition out of Band Degeneracy
  Lifting}}},\ }\href {\doibase 10.1103/PhysRevLett.102.126401} {\bibfield
  {journal} {\bibinfo  {journal} {Physical Review Letters}\ }\textbf {\bibinfo
  {volume} {102}},\ \bibinfo {pages} {126401} (\bibinfo {year}
  {2009})}\BibitemShut {NoStop}%
\bibitem [{\citenamefont {Korshunov}\ \emph {et~al.}(2014)\citenamefont
  {Korshunov}, \citenamefont {Efremov}, \citenamefont {Golubov},\ and\
  \citenamefont {Dolgov}}]{Korshunov2014}%
  \BibitemOpen
  \bibfield  {author} {\bibinfo {author} {\bibfnamefont {M.~M.}\ \bibnamefont
  {Korshunov}}, \bibinfo {author} {\bibfnamefont {D.~V.}\ \bibnamefont
  {Efremov}}, \bibinfo {author} {\bibfnamefont {A.~A.}\ \bibnamefont
  {Golubov}}, \ and\ \bibinfo {author} {\bibfnamefont {O.~V.}\ \bibnamefont
  {Dolgov}},\ }\bibfield  {title} {\emph {\bibinfo {title} {{Unexpected impact
  of magnetic disorder on multiband superconductivity}}},\ }\href {\doibase
  10.1103/PhysRevB.90.134517} {\bibfield  {journal} {\bibinfo  {journal}
  {Physical Review B}\ }\textbf {\bibinfo {volume} {90}},\ \bibinfo {pages}
  {134517} (\bibinfo {year} {2014})}\BibitemShut {NoStop}%
\bibitem [{\citenamefont {B{\"o}hmer}\ \emph {et~al.}(2014)\citenamefont
  {B{\"o}hmer}, \citenamefont {Burger}, \citenamefont {Hardy}, \citenamefont
  {Wolf}, \citenamefont {Schweiss}, \citenamefont {Fromknecht}, \citenamefont
  {Reinecker}, \citenamefont {Schranz},\ and\ \citenamefont
  {Meingast}}]{Boehmer2014}%
  \BibitemOpen
  \bibfield  {author} {\bibinfo {author} {\bibfnamefont {A.~E.}\ \bibnamefont
  {B{\"o}hmer}}, \bibinfo {author} {\bibfnamefont {P.}~\bibnamefont {Burger}},
  \bibinfo {author} {\bibfnamefont {F.}~\bibnamefont {Hardy}}, \bibinfo
  {author} {\bibfnamefont {T.}~\bibnamefont {Wolf}}, \bibinfo {author}
  {\bibfnamefont {P.}~\bibnamefont {Schweiss}}, \bibinfo {author}
  {\bibfnamefont {R.}~\bibnamefont {Fromknecht}}, \bibinfo {author}
  {\bibfnamefont {M.}~\bibnamefont {Reinecker}}, \bibinfo {author}
  {\bibfnamefont {W.}~\bibnamefont {Schranz}}, \ and\ \bibinfo {author}
  {\bibfnamefont {C.}~\bibnamefont {Meingast}},\ }\bibfield  {title} {\emph
  {\bibinfo {title} {{Nematic Susceptibility of Hole-Doped and Electron-Doped
  ${\mathrm{Ba}\mathrm{F}\mathrm{e}}_{2}{\mathrm{As}}_{2}$ Iron-Based
  Superconductors from Shear Modulus Measurements}}},\ }\href {\doibase
  10.1103/PhysRevLett.112.047001} {\bibfield  {journal} {\bibinfo  {journal}
  {Physical Review Letters}\ }\textbf {\bibinfo {volume} {112}},\ \bibinfo
  {pages} {047001} (\bibinfo {year} {2014})}\BibitemShut {NoStop}%
\bibitem [{\citenamefont {{B{\"o}hmer A. E.}}\ \emph
  {et~al.}(2015)\citenamefont {{B{\"o}hmer A. E.}}, \citenamefont {{Hardy F.}},
  \citenamefont {{Wang L.}}, \citenamefont {{Wolf T.}}, \citenamefont
  {{Schweiss P.}},\ and\ \citenamefont {{Meingast C.}}}]{Boehmer2015}%
  \BibitemOpen
  \bibfield  {author} {\bibinfo {author} {\bibnamefont {{B{\"o}hmer A. E.}}},
  \bibinfo {author} {\bibnamefont {{Hardy F.}}}, \bibinfo {author}
  {\bibnamefont {{Wang L.}}}, \bibinfo {author} {\bibnamefont {{Wolf T.}}},
  \bibinfo {author} {\bibnamefont {{Schweiss P.}}}, \ and\ \bibinfo {author}
  {\bibnamefont {{Meingast C.}}},\ }\bibfield  {title} {\emph {\bibinfo {title}
  {{Superconductivity-induced re-entrance of the orthorhombic distortion in
  Ba$_{1-x}$K$_x$Fe$_2$As$_2$}}},\ }\href {\doibase 10.1038/ncomms8911}
  {\bibfield  {journal} {\bibinfo  {journal} {Nature Communications}\ }\textbf
  {\bibinfo {volume} {6}},\ \bibinfo {pages} {7911} (\bibinfo {year}
  {2015})}\BibitemShut {NoStop}%
\bibitem [{\citenamefont {Maiti}\ and\ \citenamefont
  {Hirschfeld}(2015)}]{Maiti2015}%
  \BibitemOpen
  \bibfield  {author} {\bibinfo {author} {\bibfnamefont {S.}~\bibnamefont
  {Maiti}}\ and\ \bibinfo {author} {\bibfnamefont {P.~J.}\ \bibnamefont
  {Hirschfeld}},\ }\bibfield  {title} {\emph {\bibinfo {title} {{Collective
  modes in superconductors with competing $s$- and $d$-wave interactions}}},\
  }\href {\doibase 10.1103/PhysRevB.92.094506} {\bibfield  {journal} {\bibinfo
  {journal} {Physical Review B}\ }\textbf {\bibinfo {volume} {92}},\ \bibinfo
  {pages} {094506} (\bibinfo {year} {2015})}\BibitemShut {NoStop}%
\bibitem [{\citenamefont {Moll}\ \emph {et~al.}(2015)\citenamefont {Moll},
  \citenamefont {Zeng}, \citenamefont {Balicas}, \citenamefont {Galeski},
  \citenamefont {Balakirev}, \citenamefont {Bauer},\ and\ \citenamefont
  {Ronning}}]{Moll2015}%
  \BibitemOpen
  \bibfield  {author} {\bibinfo {author} {\bibfnamefont {P.~J.~W.}\
  \bibnamefont {Moll}}, \bibinfo {author} {\bibfnamefont {B.}~\bibnamefont
  {Zeng}}, \bibinfo {author} {\bibfnamefont {L.}~\bibnamefont {Balicas}},
  \bibinfo {author} {\bibfnamefont {S.}~\bibnamefont {Galeski}}, \bibinfo
  {author} {\bibfnamefont {F.~F.}\ \bibnamefont {Balakirev}}, \bibinfo {author}
  {\bibfnamefont {E.~D.}\ \bibnamefont {Bauer}}, \ and\ \bibinfo {author}
  {\bibfnamefont {F.}~\bibnamefont {Ronning}},\ }\bibfield  {title} {\emph
  {\bibinfo {title} {{Field-induced density wave in the heavy-fermion compound
  $\mathrm{CeRhIn}_5$}}},\ }\href {\doibase 10.1038/ncomms7663} {\bibfield
  {journal} {\bibinfo  {journal} {Nature Communications}\ }\textbf {\bibinfo
  {volume} {6}},\ \bibinfo {pages} {6663} (\bibinfo {year} {2015})}\BibitemShut
  {NoStop}%
\bibitem [{\citenamefont {Kim}\ \emph {et~al.}(2018)\citenamefont {Kim},
  \citenamefont {Souliou}, \citenamefont {Barber}, \citenamefont
  {Lefran\c{c}ois}, \citenamefont {Minola}, \citenamefont {Tortora},
  \citenamefont {Heid}, \citenamefont {Nandi}, \citenamefont {Borzi},
  \citenamefont {Garbarino}, \citenamefont {Bosak}, \citenamefont {Porras},
  \citenamefont {Loew}, \citenamefont {K{\"o}nig}, \citenamefont {Moll},
  \citenamefont {Mackenzie}, \citenamefont {Keimer}, \citenamefont {Hicks},\
  and\ \citenamefont {{Le Tacon}}}]{Kim2018}%
  \BibitemOpen
  \bibfield  {author} {\bibinfo {author} {\bibfnamefont {H.-H.}\ \bibnamefont
  {Kim}}, \bibinfo {author} {\bibfnamefont {S.~M.}\ \bibnamefont {Souliou}},
  \bibinfo {author} {\bibfnamefont {M.~E.}\ \bibnamefont {Barber}}, \bibinfo
  {author} {\bibfnamefont {E.}~\bibnamefont {Lefran\c{c}ois}}, \bibinfo
  {author} {\bibfnamefont {M.}~\bibnamefont {Minola}}, \bibinfo {author}
  {\bibfnamefont {M.}~\bibnamefont {Tortora}}, \bibinfo {author} {\bibfnamefont
  {R.}~\bibnamefont {Heid}}, \bibinfo {author} {\bibfnamefont {N.}~\bibnamefont
  {Nandi}}, \bibinfo {author} {\bibfnamefont {R.~A.}\ \bibnamefont {Borzi}},
  \bibinfo {author} {\bibfnamefont {G.}~\bibnamefont {Garbarino}}, \bibinfo
  {author} {\bibfnamefont {A.}~\bibnamefont {Bosak}}, \bibinfo {author}
  {\bibfnamefont {J.}~\bibnamefont {Porras}}, \bibinfo {author} {\bibfnamefont
  {T.}~\bibnamefont {Loew}}, \bibinfo {author} {\bibfnamefont {M.}~\bibnamefont
  {K{\"o}nig}}, \bibinfo {author} {\bibfnamefont {P.~J.~W.}\ \bibnamefont
  {Moll}}, \bibinfo {author} {\bibfnamefont {A.~P.}\ \bibnamefont {Mackenzie}},
  \bibinfo {author} {\bibfnamefont {B.}~\bibnamefont {Keimer}}, \bibinfo
  {author} {\bibfnamefont {C.~W.}\ \bibnamefont {Hicks}}, \ and\ \bibinfo
  {author} {\bibfnamefont {M.}~\bibnamefont {{Le Tacon}}},\ }\bibfield  {title}
  {\emph {\bibinfo {title} {{Uniaxial pressure control of competing orders in a
  high-temperature superconductor}}},\ }\href {\doibase
  10.1126/science.aat4708} {\bibfield  {journal} {\bibinfo  {journal}
  {Science}\ }\textbf {\bibinfo {volume} {362}},\ \bibinfo {pages} {1040}
  (\bibinfo {year} {2018})}\BibitemShut {NoStop}%
\bibitem [{\citenamefont {Maier}\ \emph {et~al.}(2019)\citenamefont {Maier},
  \citenamefont {Berlijn},\ and\ \citenamefont {Scalapino}}]{Maier2019}%
  \BibitemOpen
  \bibfield  {author} {\bibinfo {author} {\bibfnamefont {T.}~\bibnamefont
  {Maier}}, \bibinfo {author} {\bibfnamefont {T.}~\bibnamefont {Berlijn}}, \
  and\ \bibinfo {author} {\bibfnamefont {D.~J.}\ \bibnamefont {Scalapino}},\
  }\bibfield  {title} {\emph {\bibinfo {title} {{Two pairing domes as
  ${\mathrm{Cu}}^{2+}$ varies to ${\mathrm{Cu}}^{3+}$}}},\ }\href {\doibase
  10.1103/PhysRevB.99.224515} {\bibfield  {journal} {\bibinfo  {journal}
  {Physical Review B}\ }\textbf {\bibinfo {volume} {99}},\ \bibinfo {pages}
  {224515} (\bibinfo {year} {2019})}\BibitemShut {NoStop}%
\bibitem [{\citenamefont {Li}\ \emph {et~al.}(2019)\citenamefont {Li},
  \citenamefont {Ueland}, \citenamefont {Jayasekara}, \citenamefont
  {Abernathy}, \citenamefont {Sangeetha}, \citenamefont {Johnston},
  \citenamefont {Ding}, \citenamefont {Furukawa}, \citenamefont {Orth},
  \citenamefont {Kreyssig}, \citenamefont {Goldman},\ and\ \citenamefont
  {McQueeney}}]{Li2019}%
  \BibitemOpen
  \bibfield  {author} {\bibinfo {author} {\bibfnamefont {B.}~\bibnamefont
  {Li}}, \bibinfo {author} {\bibfnamefont {B.~G.}\ \bibnamefont {Ueland}},
  \bibinfo {author} {\bibfnamefont {W.~T.}\ \bibnamefont {Jayasekara}},
  \bibinfo {author} {\bibfnamefont {D.~L.}\ \bibnamefont {Abernathy}}, \bibinfo
  {author} {\bibfnamefont {N.~S.}\ \bibnamefont {Sangeetha}}, \bibinfo {author}
  {\bibfnamefont {D.~C.}\ \bibnamefont {Johnston}}, \bibinfo {author}
  {\bibfnamefont {Q.-P.}\ \bibnamefont {Ding}}, \bibinfo {author}
  {\bibfnamefont {Y.}~\bibnamefont {Furukawa}}, \bibinfo {author}
  {\bibfnamefont {P.~P.}\ \bibnamefont {Orth}}, \bibinfo {author}
  {\bibfnamefont {A.}~\bibnamefont {Kreyssig}}, \bibinfo {author}
  {\bibfnamefont {A.~I.}\ \bibnamefont {Goldman}}, \ and\ \bibinfo {author}
  {\bibfnamefont {R.~J.}\ \bibnamefont {McQueeney}},\ }\bibfield  {title}
  {\emph {\bibinfo {title} {{Competing magnetic phases and itinerant magnetic
  frustration in ${\mathrm{SrCo}}_{2}{\mathrm{As}}_{2}$}}},\ }\href {\doibase
  10.1103/PhysRevB.100.054411} {\bibfield  {journal} {\bibinfo  {journal}
  {Physical Review B}\ }\textbf {\bibinfo {volume} {100}},\ \bibinfo {pages}
  {054411} (\bibinfo {year} {2019})}\BibitemShut {NoStop}%
\bibitem [{\citenamefont {Willa}\ \emph
  {et~al.}(2019{\natexlab{a}})\citenamefont {Willa}, \citenamefont {Willa},
  \citenamefont {Bao}, \citenamefont {Koshelev}, \citenamefont {Chung},
  \citenamefont {Kanatzidis}, \citenamefont {Kwok},\ and\ \citenamefont
  {Welp}}]{WillaK2019a}%
  \BibitemOpen
  \bibfield  {author} {\bibinfo {author} {\bibfnamefont {K.}~\bibnamefont
  {Willa}}, \bibinfo {author} {\bibfnamefont {R.}~\bibnamefont {Willa}},
  \bibinfo {author} {\bibfnamefont {J.-K.}\ \bibnamefont {Bao}}, \bibinfo
  {author} {\bibfnamefont {A.~E.}\ \bibnamefont {Koshelev}}, \bibinfo {author}
  {\bibfnamefont {D.~Y.}\ \bibnamefont {Chung}}, \bibinfo {author}
  {\bibfnamefont {M.~G.}\ \bibnamefont {Kanatzidis}}, \bibinfo {author}
  {\bibfnamefont {W.-K.}\ \bibnamefont {Kwok}}, \ and\ \bibinfo {author}
  {\bibfnamefont {U.}~\bibnamefont {Welp}},\ }\bibfield  {title} {\emph
  {\bibinfo {title} {{Strongly fluctuating moments in the high-temperature
  magnetic superconductor ${\mathrm{RbEuFe}}_{4}{\mathrm{As}}_{4}$}}},\ }\href
  {\doibase 10.1103/PhysRevB.99.180502} {\bibfield  {journal} {\bibinfo
  {journal} {Physical Review B}\ }\textbf {\bibinfo {volume} {99}},\ \bibinfo
  {pages} {180502} (\bibinfo {year} {2019}{\natexlab{a}})}\BibitemShut
  {NoStop}%
\bibitem [{\citenamefont {Agterberg}\ \emph {et~al.}(2020)\citenamefont
  {Agterberg}, \citenamefont {Davis}, \citenamefont {Edkins}, \citenamefont
  {Fradkin}, \citenamefont {{Van Harlingen}}, \citenamefont {Kivelson},
  \citenamefont {Lee}, \citenamefont {Radzihovsky}, \citenamefont {Tranquada},\
  and\ \citenamefont {Wang}}]{Agterberg2020}%
  \BibitemOpen
  \bibfield  {author} {\bibinfo {author} {\bibfnamefont {D.~F.}\ \bibnamefont
  {Agterberg}}, \bibinfo {author} {\bibfnamefont {J.~S.}\ \bibnamefont
  {Davis}}, \bibinfo {author} {\bibfnamefont {S.~D.}\ \bibnamefont {Edkins}},
  \bibinfo {author} {\bibfnamefont {E.}~\bibnamefont {Fradkin}}, \bibinfo
  {author} {\bibfnamefont {D.~J.}\ \bibnamefont {{Van Harlingen}}}, \bibinfo
  {author} {\bibfnamefont {S.~A.}\ \bibnamefont {Kivelson}}, \bibinfo {author}
  {\bibfnamefont {P.~A.}\ \bibnamefont {Lee}}, \bibinfo {author} {\bibfnamefont
  {L.}~\bibnamefont {Radzihovsky}}, \bibinfo {author} {\bibfnamefont {J.~M.}\
  \bibnamefont {Tranquada}}, \ and\ \bibinfo {author} {\bibfnamefont
  {Y.}~\bibnamefont {Wang}},\ }\bibfield  {title} {\emph {\bibinfo {title}
  {{The Physics of Pair-Density Waves: Cuprate Superconductors and Beyond}}},\
  }\href {\doibase 10.1146/annurev-conmatphys-031119-050711} {\bibfield
  {journal} {\bibinfo  {journal} {Annual Review of Condensed Matter Physics}\
  }\textbf {\bibinfo {volume} {11}},\ \bibinfo {pages} {231} (\bibinfo {year}
  {2020})}\BibitemShut {NoStop}%
\bibitem [{\citenamefont {Graser}\ \emph {et~al.}(2009)\citenamefont {Graser},
  \citenamefont {Maier}, \citenamefont {Hirschfeld},\ and\ \citenamefont
  {Scalapino}}]{Graser2009}%
  \BibitemOpen
  \bibfield  {author} {\bibinfo {author} {\bibfnamefont {S.}~\bibnamefont
  {Graser}}, \bibinfo {author} {\bibfnamefont {T.~A.}\ \bibnamefont {Maier}},
  \bibinfo {author} {\bibfnamefont {P.~J.}\ \bibnamefont {Hirschfeld}}, \ and\
  \bibinfo {author} {\bibfnamefont {D.~J.}\ \bibnamefont {Scalapino}},\
  }\bibfield  {title} {\emph {\bibinfo {title} {{Near-degeneracy of several
  pairing channels in multiorbital models for the $\mathrm{Fe}$ pnictides}}},\
  }\href {\doibase 10.1088/1367-2630/11/2/025016} {\bibfield  {journal}
  {\bibinfo  {journal} {New Journal of Physics}\ }\textbf {\bibinfo {volume}
  {11}},\ \bibinfo {pages} {025016} (\bibinfo {year} {2009})}\BibitemShut
  {NoStop}%
\bibitem [{\citenamefont {{Fernandes R. M.}}\ \emph {et~al.}(2014)\citenamefont
  {{Fernandes R. M.}}, \citenamefont {{Chubukov A. V.}},\ and\ \citenamefont
  {{Schmalian J.}}}]{Fernandes2014}%
  \BibitemOpen
  \bibfield  {author} {\bibinfo {author} {\bibnamefont {{Fernandes R. M.}}},
  \bibinfo {author} {\bibnamefont {{Chubukov A. V.}}}, \ and\ \bibinfo {author}
  {\bibnamefont {{Schmalian J.}}},\ }\bibfield  {title} {\emph {\bibinfo
  {title} {{What drives nematic order in iron-based superconductors?}}},\
  }\href {\doibase 10.1038/nphys2877} {\bibfield  {journal} {\bibinfo
  {journal} {Nature Physics}\ }\textbf {\bibinfo {volume} {10}},\ \bibinfo
  {pages} {97} (\bibinfo {year} {2014})}\BibitemShut {NoStop}%
\bibitem [{\citenamefont {Wang}\ \emph {et~al.}(2019)\citenamefont {Wang},
  \citenamefont {He}, \citenamefont {Scherer}, \citenamefont {Hardy},
  \citenamefont {Schweiss}, \citenamefont {Wolf}, \citenamefont {Merz},
  \citenamefont {Andersen},\ and\ \citenamefont {Meingast}}]{Wang2019}%
  \BibitemOpen
  \bibfield  {author} {\bibinfo {author} {\bibfnamefont {L.}~\bibnamefont
  {Wang}}, \bibinfo {author} {\bibfnamefont {M.}~\bibnamefont {He}}, \bibinfo
  {author} {\bibfnamefont {D.~D.}\ \bibnamefont {Scherer}}, \bibinfo {author}
  {\bibfnamefont {F.}~\bibnamefont {Hardy}}, \bibinfo {author} {\bibfnamefont
  {P.}~\bibnamefont {Schweiss}}, \bibinfo {author} {\bibfnamefont
  {T.}~\bibnamefont {Wolf}}, \bibinfo {author} {\bibfnamefont {M.}~\bibnamefont
  {Merz}}, \bibinfo {author} {\bibfnamefont {B.~M.}\ \bibnamefont {Andersen}},
  \ and\ \bibinfo {author} {\bibfnamefont {C.}~\bibnamefont {Meingast}},\
  }\bibfield  {title} {\emph {\bibinfo {title} {{Competing Electronic Phases
  near the Onset of Superconductivity in Hole-doped
  $\mathrm{SrFe}_2\mathrm{As}_2$}}},\ }\href {\doibase 10.7566/JPSJ.88.104710}
  {\bibfield  {journal} {\bibinfo  {journal} {Journal of the Physical Society
  of Japan}\ }\textbf {\bibinfo {volume} {88}},\ \bibinfo {pages} {104710}
  (\bibinfo {year} {2019})}\BibitemShut {NoStop}%
\bibitem [{\citenamefont {Fernandes}\ \emph {et~al.}(2019)\citenamefont
  {Fernandes}, \citenamefont {Orth},\ and\ \citenamefont
  {Schmalian}}]{Fernandes2019}%
  \BibitemOpen
  \bibfield  {author} {\bibinfo {author} {\bibfnamefont {R.~M.}\ \bibnamefont
  {Fernandes}}, \bibinfo {author} {\bibfnamefont {P.~P.}\ \bibnamefont {Orth}},
  \ and\ \bibinfo {author} {\bibfnamefont {J.}~\bibnamefont {Schmalian}},\
  }\bibfield  {title} {\emph {\bibinfo {title} {{Intertwined Vestigial Order in
  Quantum Materials: Nematicity and Beyond}}},\ }\href {\doibase
  10.1146/annurev-conmatphys-031218-013200} {\bibfield  {journal} {\bibinfo
  {journal} {Annual Review of Condensed Matter Physics}\ }\textbf {\bibinfo
  {volume} {10}},\ \bibinfo {pages} {133} (\bibinfo {year} {2019})}\BibitemShut
  {NoStop}%
\bibitem [{\citenamefont {Okazaki}\ \emph {et~al.}(2011)\citenamefont
  {Okazaki}, \citenamefont {Shibauchi}, \citenamefont {Shi}, \citenamefont
  {Haga}, \citenamefont {Matsuda}, \citenamefont {Yamamoto}, \citenamefont
  {Onuki}, \citenamefont {Ikeda},\ and\ \citenamefont {Matsuda}}]{Okazaki2011}%
  \BibitemOpen
  \bibfield  {author} {\bibinfo {author} {\bibfnamefont {R.}~\bibnamefont
  {Okazaki}}, \bibinfo {author} {\bibfnamefont {T.}~\bibnamefont {Shibauchi}},
  \bibinfo {author} {\bibfnamefont {H.~J.}\ \bibnamefont {Shi}}, \bibinfo
  {author} {\bibfnamefont {Y.}~\bibnamefont {Haga}}, \bibinfo {author}
  {\bibfnamefont {T.~D.}\ \bibnamefont {Matsuda}}, \bibinfo {author}
  {\bibfnamefont {E.}~\bibnamefont {Yamamoto}}, \bibinfo {author}
  {\bibfnamefont {Y.}~\bibnamefont {Onuki}}, \bibinfo {author} {\bibfnamefont
  {H.}~\bibnamefont {Ikeda}}, \ and\ \bibinfo {author} {\bibfnamefont
  {Y.}~\bibnamefont {Matsuda}},\ }\bibfield  {title} {\emph {\bibinfo {title}
  {{Rotational Symmetry Breaking in the Hidden-Order Phase of
  $\mathrm{URu}_2\mathrm{Si}2$}}},\ }\href {\doibase 10.1126/science.1197358}
  {\bibfield  {journal} {\bibinfo  {journal} {Science}\ }\textbf {\bibinfo
  {volume} {331}},\ \bibinfo {pages} {439} (\bibinfo {year}
  {2011})}\BibitemShut {NoStop}%
\bibitem [{\citenamefont {Premala}\ \emph {et~al.}(2013)\citenamefont
  {Premala}, \citenamefont {Piers},\ and\ \citenamefont
  {Rebecca}}]{Chandra2013}%
  \BibitemOpen
  \bibfield  {author} {\bibinfo {author} {\bibfnamefont {C.}~\bibnamefont
  {Premala}}, \bibinfo {author} {\bibfnamefont {C.}~\bibnamefont {Piers}}, \
  and\ \bibinfo {author} {\bibfnamefont {F.}~\bibnamefont {Rebecca}},\
  }\bibfield  {title} {\emph {\bibinfo {title} {{Hastatic order in the
  heavy-fermion compound $\mathrm{URu}_2\mathrm{Si}_2$}}},\ }\href {\doibase
  10.1038/nature11820} {\bibfield  {journal} {\bibinfo  {journal} {Nature}\
  }\textbf {\bibinfo {volume} {493}},\ \bibinfo {pages} {621} (\bibinfo {year}
  {2013})}\BibitemShut {NoStop}%
\bibitem [{\citenamefont {Kung}\ \emph {et~al.}(2015)\citenamefont {Kung},
  \citenamefont {Baumbach}, \citenamefont {Bauer}, \citenamefont
  {Thorsm{\o}lle}, \citenamefont {Zhang}, \citenamefont {Haule}, \citenamefont
  {Mydosh},\ and\ \citenamefont {Blumberg}}]{Kung2015}%
  \BibitemOpen
  \bibfield  {author} {\bibinfo {author} {\bibfnamefont {H.-H.}\ \bibnamefont
  {Kung}}, \bibinfo {author} {\bibfnamefont {R.~E.}\ \bibnamefont {Baumbach}},
  \bibinfo {author} {\bibfnamefont {E.~D.}\ \bibnamefont {Bauer}}, \bibinfo
  {author} {\bibfnamefont {V.~K.}\ \bibnamefont {Thorsm{\o}lle}}, \bibinfo
  {author} {\bibfnamefont {W.-L.}\ \bibnamefont {Zhang}}, \bibinfo {author}
  {\bibfnamefont {K.}~\bibnamefont {Haule}}, \bibinfo {author} {\bibfnamefont
  {J.~A.}\ \bibnamefont {Mydosh}}, \ and\ \bibinfo {author} {\bibfnamefont
  {G.}~\bibnamefont {Blumberg}},\ }\bibfield  {title} {\emph {\bibinfo {title}
  {{Chirality density wave of the 'hidden order' phase in
  $\mathrm{URu}_2\mathrm{Si}_2$}}},\ }\href {\doibase 10.1126/science.1259729}
  {\bibfield  {journal} {\bibinfo  {journal} {Science}\ }\textbf {\bibinfo
  {volume} {347}},\ \bibinfo {pages} {1339} (\bibinfo {year}
  {2015})}\BibitemShut {NoStop}%
\bibitem [{\citenamefont {Emery}\ and\ \citenamefont
  {Kivelson}(1995)}]{Emery1995}%
  \BibitemOpen
  \bibfield  {author} {\bibinfo {author} {\bibfnamefont {V.~J.}\ \bibnamefont
  {Emery}}\ and\ \bibinfo {author} {\bibfnamefont {S.~A.}\ \bibnamefont
  {Kivelson}},\ }\bibfield  {title} {\emph {\bibinfo {title} {{Importance of
  phase fluctuations in superconductors with small superfluid density}}},\
  }\href@noop {} {\bibfield  {journal} {\bibinfo  {journal} {Nature}\ }\textbf
  {\bibinfo {volume} {374}},\ \bibinfo {pages} {434} (\bibinfo {year}
  {1995})}\BibitemShut {NoStop}%
\bibitem [{\citenamefont {Chakravarty}\ \emph {et~al.}(2001)\citenamefont
  {Chakravarty}, \citenamefont {Laughlin}, \citenamefont {Morr},\ and\
  \citenamefont {Nayak}}]{Chakravarty2001}%
  \BibitemOpen
  \bibfield  {author} {\bibinfo {author} {\bibfnamefont {S.}~\bibnamefont
  {Chakravarty}}, \bibinfo {author} {\bibfnamefont {R.~B.}\ \bibnamefont
  {Laughlin}}, \bibinfo {author} {\bibfnamefont {D.~K.}\ \bibnamefont {Morr}},
  \ and\ \bibinfo {author} {\bibfnamefont {C.}~\bibnamefont {Nayak}},\
  }\bibfield  {title} {\emph {\bibinfo {title} {{Hidden order in the
  cuprates}}},\ }\href {\doibase 10.1103/PhysRevB.63.094503} {\bibfield
  {journal} {\bibinfo  {journal} {Physical Review B}\ }\textbf {\bibinfo
  {volume} {63}},\ \bibinfo {pages} {094503} (\bibinfo {year}
  {2001})}\BibitemShut {NoStop}%
\bibitem [{\citenamefont {Raghu}\ \emph {et~al.}(2010)\citenamefont {Raghu},
  \citenamefont {Kapitulnik},\ and\ \citenamefont {Kivelson}}]{Raghu2010}%
  \BibitemOpen
  \bibfield  {author} {\bibinfo {author} {\bibfnamefont {S.}~\bibnamefont
  {Raghu}}, \bibinfo {author} {\bibfnamefont {A.}~\bibnamefont {Kapitulnik}}, \
  and\ \bibinfo {author} {\bibfnamefont {S.~A.}\ \bibnamefont {Kivelson}},\
  }\bibfield  {title} {\emph {\bibinfo {title} {{Hidden Quasi-One-Dimensional
  Superconductivity in ${\mathrm{Sr}}_{2}{\mathrm{RuO}}_{4}$}}},\ }\href
  {\doibase 10.1103/PhysRevLett.105.136401} {\bibfield  {journal} {\bibinfo
  {journal} {Physical Review Letters}\ }\textbf {\bibinfo {volume} {105}},\
  \bibinfo {pages} {136401} (\bibinfo {year} {2010})}\BibitemShut {NoStop}%
\bibitem [{\citenamefont {Scaffidi}\ \emph {et~al.}(2014)\citenamefont
  {Scaffidi}, \citenamefont {Romers},\ and\ \citenamefont
  {Simon}}]{Scaffidi2014}%
  \BibitemOpen
  \bibfield  {author} {\bibinfo {author} {\bibfnamefont {T.}~\bibnamefont
  {Scaffidi}}, \bibinfo {author} {\bibfnamefont {J.~C.}\ \bibnamefont
  {Romers}}, \ and\ \bibinfo {author} {\bibfnamefont {S.~H.}\ \bibnamefont
  {Simon}},\ }\bibfield  {title} {\emph {\bibinfo {title} {{Pairing symmetry
  and dominant band in ${\mathrm{Sr}}_{2}{\mathrm{RuO}}_{4}$}}},\ }\href
  {\doibase 10.1103/PhysRevB.89.220510} {\bibfield  {journal} {\bibinfo
  {journal} {Physical Review B}\ }\textbf {\bibinfo {volume} {89}},\ \bibinfo
  {pages} {220510} (\bibinfo {year} {2014})}\BibitemShut {NoStop}%
\bibitem [{\citenamefont {Steppke}\ \emph {et~al.}(2017)\citenamefont
  {Steppke}, \citenamefont {Zhao}, \citenamefont {Barber}, \citenamefont
  {Scaffidi}, \citenamefont {Jerzembeck}, \citenamefont {Rosner}, \citenamefont
  {Gibbs}, \citenamefont {Maeno}, \citenamefont {Simon}, \citenamefont
  {Mackenzie},\ and\ \citenamefont {Hicks}}]{Steppke2017}%
  \BibitemOpen
  \bibfield  {author} {\bibinfo {author} {\bibfnamefont {A.}~\bibnamefont
  {Steppke}}, \bibinfo {author} {\bibfnamefont {L.}~\bibnamefont {Zhao}},
  \bibinfo {author} {\bibfnamefont {M.~E.}\ \bibnamefont {Barber}}, \bibinfo
  {author} {\bibfnamefont {T.}~\bibnamefont {Scaffidi}}, \bibinfo {author}
  {\bibfnamefont {F.}~\bibnamefont {Jerzembeck}}, \bibinfo {author}
  {\bibfnamefont {H.}~\bibnamefont {Rosner}}, \bibinfo {author} {\bibfnamefont
  {A.~S.}\ \bibnamefont {Gibbs}}, \bibinfo {author} {\bibfnamefont
  {Y.}~\bibnamefont {Maeno}}, \bibinfo {author} {\bibfnamefont {S.~H.}\
  \bibnamefont {Simon}}, \bibinfo {author} {\bibfnamefont {A.~P.}\ \bibnamefont
  {Mackenzie}}, \ and\ \bibinfo {author} {\bibfnamefont {C.~W.}\ \bibnamefont
  {Hicks}},\ }\bibfield  {title} {\emph {\bibinfo {title} {{Strong peak in
  $T_c$ of $\mathrm{Sr}_2\mathrm{Ru}\mathrm{O}_4$ under uniaxial pressure}}},\
  }\href {\doibase 10.1126/science.aaf9398} {\bibfield  {journal} {\bibinfo
  {journal} {Science}\ }\textbf {\bibinfo {volume} {355}} (\bibinfo {year}
  {2017}),\ 10.1126/science.aaf9398}\BibitemShut {NoStop}%
\bibitem [{\citenamefont {Zhao}\ \emph {et~al.}(2015)\citenamefont {Zhao},
  \citenamefont {Torchinsky}, \citenamefont {Chu}, \citenamefont {Ivanov},
  \citenamefont {Lifshitz}, \citenamefont {Flint}, \citenamefont {Qi},
  \citenamefont {Cao},\ and\ \citenamefont {Hsieh}}]{Zhao2015}%
  \BibitemOpen
  \bibfield  {author} {\bibinfo {author} {\bibfnamefont {L.}~\bibnamefont
  {Zhao}}, \bibinfo {author} {\bibfnamefont {D.~H.}\ \bibnamefont
  {Torchinsky}}, \bibinfo {author} {\bibfnamefont {H.}~\bibnamefont {Chu}},
  \bibinfo {author} {\bibfnamefont {V.}~\bibnamefont {Ivanov}}, \bibinfo
  {author} {\bibfnamefont {R.}~\bibnamefont {Lifshitz}}, \bibinfo {author}
  {\bibfnamefont {R.}~\bibnamefont {Flint}}, \bibinfo {author} {\bibfnamefont
  {T.}~\bibnamefont {Qi}}, \bibinfo {author} {\bibfnamefont {G.}~\bibnamefont
  {Cao}}, \ and\ \bibinfo {author} {\bibfnamefont {D.}~\bibnamefont {Hsieh}},\
  }\bibfield  {title} {\emph {\bibinfo {title} {{Evidence of an odd-parity
  hidden order in a spin--orbit coupled correlated iridate}}},\ }\href
  {https://doi.org/10.1038/nphys3517} {\bibfield  {journal} {\bibinfo
  {journal} {Nature Physics}\ }\textbf {\bibinfo {volume} {12}},\ \bibinfo
  {pages} {32} (\bibinfo {year} {2015})}\BibitemShut {NoStop}%
\bibitem [{\citenamefont {Fechner}\ \emph {et~al.}(2016)\citenamefont
  {Fechner}, \citenamefont {Fierz}, \citenamefont {Th{\"o}le}, \citenamefont
  {Staub},\ and\ \citenamefont {Spaldin}}]{Fechner2016}%
  \BibitemOpen
  \bibfield  {author} {\bibinfo {author} {\bibfnamefont {M.}~\bibnamefont
  {Fechner}}, \bibinfo {author} {\bibfnamefont {M.~J.~A.}\ \bibnamefont
  {Fierz}}, \bibinfo {author} {\bibfnamefont {F.}~\bibnamefont {Th{\"o}le}},
  \bibinfo {author} {\bibfnamefont {U.}~\bibnamefont {Staub}}, \ and\ \bibinfo
  {author} {\bibfnamefont {N.~A.}\ \bibnamefont {Spaldin}},\ }\bibfield
  {title} {\emph {\bibinfo {title} {{Quasistatic magnetoelectric multipoles as
  order parameter for pseudogap phase in cuprate superconductors}}},\ }\href
  {\doibase 10.1103/PhysRevB.93.174419} {\bibfield  {journal} {\bibinfo
  {journal} {Physical Review B}\ }\textbf {\bibinfo {volume} {93}},\ \bibinfo
  {pages} {174419} (\bibinfo {year} {2016})}\BibitemShut {NoStop}%
\bibitem [{\citenamefont {Pustogow}\ \emph {et~al.}(2019)\citenamefont
  {Pustogow}, \citenamefont {Luo}, \citenamefont {Chronister}, \citenamefont
  {Su}, \citenamefont {Sokolov}, \citenamefont {Jerzembeck}, \citenamefont
  {Mackenzie}, \citenamefont {Hicks}, \citenamefont {Kikugawa}, \citenamefont
  {Raghu}, \citenamefont {Bauer},\ and\ \citenamefont {Brown}}]{Pustogow2019}%
  \BibitemOpen
  \bibfield  {author} {\bibinfo {author} {\bibfnamefont {A.}~\bibnamefont
  {Pustogow}}, \bibinfo {author} {\bibfnamefont {Y.}~\bibnamefont {Luo}},
  \bibinfo {author} {\bibfnamefont {A.}~\bibnamefont {Chronister}}, \bibinfo
  {author} {\bibfnamefont {Y.~S.}\ \bibnamefont {Su}}, \bibinfo {author}
  {\bibfnamefont {D.~A.}\ \bibnamefont {Sokolov}}, \bibinfo {author}
  {\bibfnamefont {F.}~\bibnamefont {Jerzembeck}}, \bibinfo {author}
  {\bibfnamefont {A.~P.}\ \bibnamefont {Mackenzie}}, \bibinfo {author}
  {\bibfnamefont {C.~W.}\ \bibnamefont {Hicks}}, \bibinfo {author}
  {\bibfnamefont {N.}~\bibnamefont {Kikugawa}}, \bibinfo {author}
  {\bibfnamefont {S.}~\bibnamefont {Raghu}}, \bibinfo {author} {\bibfnamefont
  {E.~D.}\ \bibnamefont {Bauer}}, \ and\ \bibinfo {author} {\bibfnamefont
  {S.~E.}\ \bibnamefont {Brown}},\ }\bibfield  {title} {\emph {\bibinfo {title}
  {{Constraints on the superconducting order parameter in
  $\mathrm{Sr}_2\mathrm{Ru}\mathrm{O}_4$ from oxygen-17 nuclear magnetic
  resonance}}},\ }\href {\doibase 10.1038/s41586-019-1596-2} {\bibfield
  {journal} {\bibinfo  {journal} {Nature}\ }\textbf {\bibinfo {volume} {574}},\
  \bibinfo {pages} {72} (\bibinfo {year} {2019})}\BibitemShut {NoStop}%
\bibitem [{\citenamefont {Ramires}\ and\ \citenamefont
  {Sigrist}(2019)}]{Ramires2019}%
  \BibitemOpen
  \bibfield  {author} {\bibinfo {author} {\bibfnamefont {A.}~\bibnamefont
  {Ramires}}\ and\ \bibinfo {author} {\bibfnamefont {M.}~\bibnamefont
  {Sigrist}},\ }\bibfield  {title} {\emph {\bibinfo {title} {{Superconducting
  order parameter of ${\mathrm{Sr}}_{2}{\mathrm{RuO}}_{4}$: A microscopic
  perspective}}},\ }\href {\doibase 10.1103/PhysRevB.100.104501} {\bibfield
  {journal} {\bibinfo  {journal} {Physical Review B}\ }\textbf {\bibinfo
  {volume} {100}},\ \bibinfo {pages} {104501} (\bibinfo {year}
  {2019})}\BibitemShut {NoStop}%
\bibitem [{\citenamefont {R{\o}ising}\ \emph {et~al.}(2019)\citenamefont
  {R{\o}ising}, \citenamefont {Scaffidi}, \citenamefont {Flicker},
  \citenamefont {Lange},\ and\ \citenamefont {Simon}}]{Roising2019}%
  \BibitemOpen
  \bibfield  {author} {\bibinfo {author} {\bibfnamefont {H.~S.}\ \bibnamefont
  {R{\o}ising}}, \bibinfo {author} {\bibfnamefont {T.}~\bibnamefont
  {Scaffidi}}, \bibinfo {author} {\bibfnamefont {F.}~\bibnamefont {Flicker}},
  \bibinfo {author} {\bibfnamefont {G.~F.}\ \bibnamefont {Lange}}, \ and\
  \bibinfo {author} {\bibfnamefont {S.~H.}\ \bibnamefont {Simon}},\ }\bibfield
  {title} {\emph {\bibinfo {title} {{Superconducting order of
  ${\mathrm{Sr}}_{2}{\mathrm{RuO}}_{4}$ from a three-dimensional microscopic
  model}}},\ }\href {\doibase 10.1103/PhysRevResearch.1.033108} {\bibfield
  {journal} {\bibinfo  {journal} {Physical Review Research}\ }\textbf {\bibinfo
  {volume} {1}},\ \bibinfo {pages} {033108} (\bibinfo {year}
  {2019})}\BibitemShut {NoStop}%
\bibitem [{\citenamefont {Mackenzie}(2020)}]{Mackenzie2020}%
  \BibitemOpen
  \bibfield  {author} {\bibinfo {author} {\bibfnamefont {A.~P.}\ \bibnamefont
  {Mackenzie}},\ }\bibfield  {title} {\emph {\bibinfo {title} {{A Personal
  Perspective on the Unconventional Superconductivity of
  $\mathrm{Sr}_2\mathrm{RuO}_4$}}},\ }\href {\doibase
  10.1007/s10948-019-05312-4} {\bibfield  {journal} {\bibinfo  {journal}
  {Journal of Superconductivity and Novel Magnetism}\ }\textbf {\bibinfo
  {volume} {33}},\ \bibinfo {pages} {177} (\bibinfo {year} {2020})}\BibitemShut
  {NoStop}%
\bibitem [{\citenamefont {Luke}\ \emph {et~al.}(1998)\citenamefont {Luke},
  \citenamefont {Fudamoto}, \citenamefont {Kojima}, \citenamefont {Larkin},
  \citenamefont {Merrin}, \citenamefont {Nachumi}, \citenamefont {Uemura},
  \citenamefont {Maeno}, \citenamefont {Mao}, \citenamefont {Mori},
  \citenamefont {Nakamura},\ and\ \citenamefont {Sigrist}}]{Luke1998}%
  \BibitemOpen
  \bibfield  {author} {\bibinfo {author} {\bibfnamefont {G.~M.}\ \bibnamefont
  {Luke}}, \bibinfo {author} {\bibfnamefont {Y.}~\bibnamefont {Fudamoto}},
  \bibinfo {author} {\bibfnamefont {K.~M.}\ \bibnamefont {Kojima}}, \bibinfo
  {author} {\bibfnamefont {M.~I.}\ \bibnamefont {Larkin}}, \bibinfo {author}
  {\bibfnamefont {J.}~\bibnamefont {Merrin}}, \bibinfo {author} {\bibfnamefont
  {B.}~\bibnamefont {Nachumi}}, \bibinfo {author} {\bibfnamefont {Y.~J.}\
  \bibnamefont {Uemura}}, \bibinfo {author} {\bibfnamefont {Y.}~\bibnamefont
  {Maeno}}, \bibinfo {author} {\bibfnamefont {Z.~Q.}\ \bibnamefont {Mao}},
  \bibinfo {author} {\bibfnamefont {Y.}~\bibnamefont {Mori}}, \bibinfo {author}
  {\bibfnamefont {H.}~\bibnamefont {Nakamura}}, \ and\ \bibinfo {author}
  {\bibfnamefont {M.}~\bibnamefont {Sigrist}},\ }\bibfield  {title} {\emph
  {\bibinfo {title} {{Time-reversal symmetry-breaking superconductivity in
  $\mathrm{Sr}_2\mathrm{Ru}\mathrm{O}_4$}}},\ }\href {\doibase 10.1038/29038}
  {\bibfield  {journal} {\bibinfo  {journal} {Nature}\ }\textbf {\bibinfo
  {volume} {394}},\ \bibinfo {pages} {558} (\bibinfo {year}
  {1998})}\BibitemShut {NoStop}%
\bibitem [{\citenamefont {Kivelson}\ \emph {et~al.}(2020)\citenamefont
  {Kivelson}, \citenamefont {Yuan}, \citenamefont {Ramshaw},\ and\
  \citenamefont {Thomale}}]{Kivelson2020-condmat}%
  \BibitemOpen
  \bibfield  {author} {\bibinfo {author} {\bibfnamefont {S.~A.}\ \bibnamefont
  {Kivelson}}, \bibinfo {author} {\bibfnamefont {A.~C.}\ \bibnamefont {Yuan}},
  \bibinfo {author} {\bibfnamefont {B.~J.}\ \bibnamefont {Ramshaw}}, \ and\
  \bibinfo {author} {\bibfnamefont {R.}~\bibnamefont {Thomale}},\ }\bibfield
  {title} {\emph {\bibinfo {title} {{A proposal for reconciling diverse
  experiments on the superconducting state in
  $\mathrm{Sr}_2\mathrm{Ru}\mathrm{O}_4$}}},\ }\href@noop {} {\bibfield
  {journal} {\bibinfo  {journal} {arXiv:2002.00016 [cond-mat.supr-con]}\ }
  (\bibinfo {year} {2020})}\BibitemShut {NoStop}%
\bibitem [{\citenamefont {Ghosh}\ \emph {et~al.}(2020)\citenamefont {Ghosh},
  \citenamefont {Shekhter}, \citenamefont {Jerzembeck}, \citenamefont
  {Kikugawa}, \citenamefont {Sokolov}, \citenamefont {Brando}, \citenamefont
  {Mackenzie}, \citenamefont {Hicks},\ and\ \citenamefont
  {Ramshaw}}]{Ghosh2020-condmat}%
  \BibitemOpen
  \bibfield  {author} {\bibinfo {author} {\bibfnamefont {S.}~\bibnamefont
  {Ghosh}}, \bibinfo {author} {\bibfnamefont {A.}~\bibnamefont {Shekhter}},
  \bibinfo {author} {\bibfnamefont {F.}~\bibnamefont {Jerzembeck}}, \bibinfo
  {author} {\bibfnamefont {N.}~\bibnamefont {Kikugawa}}, \bibinfo {author}
  {\bibfnamefont {D.~A.}\ \bibnamefont {Sokolov}}, \bibinfo {author}
  {\bibfnamefont {M.}~\bibnamefont {Brando}}, \bibinfo {author} {\bibfnamefont
  {A.~P.}\ \bibnamefont {Mackenzie}}, \bibinfo {author} {\bibfnamefont {C.~W.}\
  \bibnamefont {Hicks}}, \ and\ \bibinfo {author} {\bibfnamefont {B.~J.}\
  \bibnamefont {Ramshaw}},\ }\bibfield  {title} {\emph {\bibinfo {title}
  {{Thermodynamic Evidence for a Two-Component Superconducting Order Parameter
  in $\mathrm{Sr}_2\mathrm{Ru}\mathrm{O}_4$}}},\ }\href@noop {} {\bibfield
  {journal} {\bibinfo  {journal} {arXiv:2002.06130 [cond-mat.supr-con]}\ }
  (\bibinfo {year} {2020})}\BibitemShut {NoStop}%
\bibitem [{\citenamefont {Hicks}\ \emph {et~al.}(2014)\citenamefont {Hicks},
  \citenamefont {Brodsky}, \citenamefont {Yelland}, \citenamefont {Gibbs},
  \citenamefont {Bruin}, \citenamefont {Barber}, \citenamefont {Edkins},
  \citenamefont {Nishimura}, \citenamefont {Yonezawa}, \citenamefont {Maeno},\
  and\ \citenamefont {Mackenzie}}]{Hicks2014}%
  \BibitemOpen
  \bibfield  {author} {\bibinfo {author} {\bibfnamefont {C.~W.}\ \bibnamefont
  {Hicks}}, \bibinfo {author} {\bibfnamefont {D.~O.}\ \bibnamefont {Brodsky}},
  \bibinfo {author} {\bibfnamefont {E.~A.}\ \bibnamefont {Yelland}}, \bibinfo
  {author} {\bibfnamefont {A.~S.}\ \bibnamefont {Gibbs}}, \bibinfo {author}
  {\bibfnamefont {J.~A.~N.}\ \bibnamefont {Bruin}}, \bibinfo {author}
  {\bibfnamefont {M.~E.}\ \bibnamefont {Barber}}, \bibinfo {author}
  {\bibfnamefont {S.~D.}\ \bibnamefont {Edkins}}, \bibinfo {author}
  {\bibfnamefont {K.}~\bibnamefont {Nishimura}}, \bibinfo {author}
  {\bibfnamefont {S.}~\bibnamefont {Yonezawa}}, \bibinfo {author}
  {\bibfnamefont {Y.}~\bibnamefont {Maeno}}, \ and\ \bibinfo {author}
  {\bibfnamefont {A.~P.}\ \bibnamefont {Mackenzie}},\ }\bibfield  {title}
  {\emph {\bibinfo {title} {{Strong Increase of $T_c$ of
  $\mathrm{Sr}_2\mathrm{Ru}\mathrm{O}_4$ Under Both Tensile and Compressive
  Strain}}},\ }\href {\doibase 10.1126/science.1248292} {\bibfield  {journal}
  {\bibinfo  {journal} {Science}\ }\textbf {\bibinfo {volume} {344}},\ \bibinfo
  {pages} {283} (\bibinfo {year} {2014})}\BibitemShut {NoStop}%
\bibitem [{\citenamefont {Watson}\ \emph {et~al.}(2018)\citenamefont {Watson},
  \citenamefont {Gibbs}, \citenamefont {Mackenzie}, \citenamefont {Hicks},\
  and\ \citenamefont {Moler}}]{Watson2018}%
  \BibitemOpen
  \bibfield  {author} {\bibinfo {author} {\bibfnamefont {C.~A.}\ \bibnamefont
  {Watson}}, \bibinfo {author} {\bibfnamefont {A.~S.}\ \bibnamefont {Gibbs}},
  \bibinfo {author} {\bibfnamefont {A.~P.}\ \bibnamefont {Mackenzie}}, \bibinfo
  {author} {\bibfnamefont {C.~W.}\ \bibnamefont {Hicks}}, \ and\ \bibinfo
  {author} {\bibfnamefont {K.~A.}\ \bibnamefont {Moler}},\ }\bibfield  {title}
  {\emph {\bibinfo {title} {{Micron-scale measurements of low anisotropic
  strain response of local ${T}_{c}$ in
  ${\mathrm{Sr}}_{2}{\mathrm{RuO}}_{4}$}}},\ }\href {\doibase
  10.1103/PhysRevB.98.094521} {\bibfield  {journal} {\bibinfo  {journal}
  {Physical Review B}\ }\textbf {\bibinfo {volume} {98}},\ \bibinfo {pages}
  {094521} (\bibinfo {year} {2018})}\BibitemShut {NoStop}%
\bibitem [{\citenamefont {Grinenko}\ \emph {et~al.}(2020)\citenamefont
  {Grinenko} \emph {et~al.}}]{Grinenko2020-condmat}%
  \BibitemOpen
  \bibfield  {author} {\bibinfo {author} {\bibfnamefont {V.}~\bibnamefont
  {Grinenko}} \emph {et~al.},\ }\bibfield  {title} {\emph {\bibinfo {title}
  {{Split superconducting and time-reversal symmetry-breaking transitions, and
  magnetic order in $\mathrm{Sr}_2\mathrm{Ru}\mathrm{O}_4$ under uniaxial
  stress}}},\ }\href@noop {} {\bibfield  {journal} {\bibinfo  {journal}
  {arXiv:2001.08152 [cond-mat.supr-con]}\ } (\bibinfo {year}
  {2020})}\BibitemShut {NoStop}%
\bibitem [{\citenamefont {Fang}\ \emph {et~al.}(2008)\citenamefont {Fang},
  \citenamefont {Yao}, \citenamefont {Tsai}, \citenamefont {Hu},\ and\
  \citenamefont {Kivelson}}]{Fang2008}%
  \BibitemOpen
  \bibfield  {author} {\bibinfo {author} {\bibfnamefont {C.}~\bibnamefont
  {Fang}}, \bibinfo {author} {\bibfnamefont {H.}~\bibnamefont {Yao}}, \bibinfo
  {author} {\bibfnamefont {W.-F.}\ \bibnamefont {Tsai}}, \bibinfo {author}
  {\bibfnamefont {J.}~\bibnamefont {Hu}}, \ and\ \bibinfo {author}
  {\bibfnamefont {S.~A.}\ \bibnamefont {Kivelson}},\ }\bibfield  {title} {\emph
  {\bibinfo {title} {{Theory of electron nematic order in LaFeAsO}}},\ }\href
  {\doibase 10.1103/PhysRevB.77.224509} {\bibfield  {journal} {\bibinfo
  {journal} {Phys. Rev. B}\ }\textbf {\bibinfo {volume} {77}},\ \bibinfo
  {pages} {224509} (\bibinfo {year} {2008})}\BibitemShut {NoStop}%
\bibitem [{\citenamefont {Xu}\ \emph {et~al.}(2008)\citenamefont {Xu},
  \citenamefont {M{\"u}ller},\ and\ \citenamefont {Sachdev}}]{Xu2008}%
  \BibitemOpen
  \bibfield  {author} {\bibinfo {author} {\bibfnamefont {C.}~\bibnamefont
  {Xu}}, \bibinfo {author} {\bibfnamefont {M.}~\bibnamefont {M{\"u}ller}}, \
  and\ \bibinfo {author} {\bibfnamefont {S.}~\bibnamefont {Sachdev}},\
  }\bibfield  {title} {\emph {\bibinfo {title} {{Ising and spin orders in the
  iron-based superconductors}}},\ }\href {\doibase 10.1103/PhysRevB.78.020501}
  {\bibfield  {journal} {\bibinfo  {journal} {Physical Review B}\ }\textbf
  {\bibinfo {volume} {78}},\ \bibinfo {pages} {020501} (\bibinfo {year}
  {2008})}\BibitemShut {NoStop}%
\bibitem [{\citenamefont {Fernandes}\ \emph {et~al.}(2010)\citenamefont
  {Fernandes}, \citenamefont {VanBebber}, \citenamefont {Bhattacharya},
  \citenamefont {Chandra}, \citenamefont {Keppens}, \citenamefont {Mandrus},
  \citenamefont {McGuire}, \citenamefont {Sales}, \citenamefont {Sefat},\ and\
  \citenamefont {Schmalian}}]{Fernandes2010}%
  \BibitemOpen
  \bibfield  {author} {\bibinfo {author} {\bibfnamefont {R.~M.}\ \bibnamefont
  {Fernandes}}, \bibinfo {author} {\bibfnamefont {L.~H.}\ \bibnamefont
  {VanBebber}}, \bibinfo {author} {\bibfnamefont {S.}~\bibnamefont
  {Bhattacharya}}, \bibinfo {author} {\bibfnamefont {P.}~\bibnamefont
  {Chandra}}, \bibinfo {author} {\bibfnamefont {V.}~\bibnamefont {Keppens}},
  \bibinfo {author} {\bibfnamefont {D.}~\bibnamefont {Mandrus}}, \bibinfo
  {author} {\bibfnamefont {M.~A.}\ \bibnamefont {McGuire}}, \bibinfo {author}
  {\bibfnamefont {B.~C.}\ \bibnamefont {Sales}}, \bibinfo {author}
  {\bibfnamefont {A.~S.}\ \bibnamefont {Sefat}}, \ and\ \bibinfo {author}
  {\bibfnamefont {J.}~\bibnamefont {Schmalian}},\ }\bibfield  {title} {\emph
  {\bibinfo {title} {{Effects of Nematic Fluctuations on the Elastic Properties
  of Iron Arsenide Superconductors}}},\ }\href {\doibase
  10.1103/PhysRevLett.105.157003} {\bibfield  {journal} {\bibinfo  {journal}
  {Physical Review Letters}\ }\textbf {\bibinfo {volume} {105}},\ \bibinfo
  {pages} {157003} (\bibinfo {year} {2010})}\BibitemShut {NoStop}%
\bibitem [{\citenamefont {Cano}\ \emph {et~al.}(2010)\citenamefont {Cano},
  \citenamefont {Civelli}, \citenamefont {Eremin},\ and\ \citenamefont
  {Paul}}]{Cano2010}%
  \BibitemOpen
  \bibfield  {author} {\bibinfo {author} {\bibfnamefont {A.}~\bibnamefont
  {Cano}}, \bibinfo {author} {\bibfnamefont {M.}~\bibnamefont {Civelli}},
  \bibinfo {author} {\bibfnamefont {I.}~\bibnamefont {Eremin}}, \ and\ \bibinfo
  {author} {\bibfnamefont {I.}~\bibnamefont {Paul}},\ }\bibfield  {title}
  {\emph {\bibinfo {title} {{Interplay of magnetic and structural transitions
  in iron-based pnictide superconductors}}},\ }\href {\doibase
  10.1103/PhysRevB.82.020408} {\bibfield  {journal} {\bibinfo  {journal}
  {Physical Review B}\ }\textbf {\bibinfo {volume} {82}},\ \bibinfo {pages}
  {020408} (\bibinfo {year} {2010})}\BibitemShut {NoStop}%
\bibitem [{\citenamefont {Fernandes}\ \emph {et~al.}(2012)\citenamefont
  {Fernandes}, \citenamefont {Chubukov}, \citenamefont {Knolle}, \citenamefont
  {Eremin},\ and\ \citenamefont {Schmalian}}]{Fernandes2012a}%
  \BibitemOpen
  \bibfield  {author} {\bibinfo {author} {\bibfnamefont {R.~M.}\ \bibnamefont
  {Fernandes}}, \bibinfo {author} {\bibfnamefont {A.~V.}\ \bibnamefont
  {Chubukov}}, \bibinfo {author} {\bibfnamefont {J.}~\bibnamefont {Knolle}},
  \bibinfo {author} {\bibfnamefont {I.}~\bibnamefont {Eremin}}, \ and\ \bibinfo
  {author} {\bibfnamefont {J.}~\bibnamefont {Schmalian}},\ }\bibfield  {title}
  {\emph {\bibinfo {title} {{Preemptive nematic order, pseudogap, and orbital
  order in the iron pnictides}}},\ }\href {\doibase 10.1103/PhysRevB.85.024534}
  {\bibfield  {journal} {\bibinfo  {journal} {Physical Review B}\ }\textbf
  {\bibinfo {volume} {85}},\ \bibinfo {pages} {024534} (\bibinfo {year}
  {2012})}\BibitemShut {NoStop}%
\bibitem [{\citenamefont {Fernandes}\ and\ \citenamefont
  {Schmalian}(2012)}]{Fernandes2012b}%
  \BibitemOpen
  \bibfield  {author} {\bibinfo {author} {\bibfnamefont {R.~M.}\ \bibnamefont
  {Fernandes}}\ and\ \bibinfo {author} {\bibfnamefont {J.}~\bibnamefont
  {Schmalian}},\ }\bibfield  {title} {\emph {\bibinfo {title} {{Manifestations
  of nematic degrees of freedom in the magnetic, elastic, and superconducting
  properties of the iron pnictides}}},\ }\href {\doibase
  10.1088/0953-2048/25/8/084005} {\bibfield  {journal} {\bibinfo  {journal}
  {Superconductor Science and Technology}\ }\textbf {\bibinfo {volume} {25}},\
  \bibinfo {pages} {084005} (\bibinfo {year} {2012})}\BibitemShut {NoStop}%
\bibitem [{\citenamefont {Stanev}\ and\ \citenamefont
  {Littlewood}(2013)}]{Stanev2013}%
  \BibitemOpen
  \bibfield  {author} {\bibinfo {author} {\bibfnamefont {V.}~\bibnamefont
  {Stanev}}\ and\ \bibinfo {author} {\bibfnamefont {P.~B.}\ \bibnamefont
  {Littlewood}},\ }\bibfield  {title} {\emph {\bibinfo {title} {{Nematicity
  driven by hybridization in iron-based superconductors}}},\ }\href {\doibase
  10.1103/PhysRevB.87.161122} {\bibfield  {journal} {\bibinfo  {journal}
  {Physical Review B}\ }\textbf {\bibinfo {volume} {87}},\ \bibinfo {pages}
  {161122} (\bibinfo {year} {2013})}\BibitemShut {NoStop}%
\bibitem [{\citenamefont {Willa}\ \emph
  {et~al.}(2019{\natexlab{b}})\citenamefont {Willa}, \citenamefont {Fritz},\
  and\ \citenamefont {Schmalian}}]{Willa2019a}%
  \BibitemOpen
  \bibfield  {author} {\bibinfo {author} {\bibfnamefont {R.}~\bibnamefont
  {Willa}}, \bibinfo {author} {\bibfnamefont {M.}~\bibnamefont {Fritz}}, \ and\
  \bibinfo {author} {\bibfnamefont {J.}~\bibnamefont {Schmalian}},\ }\bibfield
  {title} {\emph {\bibinfo {title} {{Strain tuning and anisotropic spin
  correlations in iron-based systems}}},\ }\href {\doibase
  10.1103/PhysRevB.100.085106} {\bibfield  {journal} {\bibinfo  {journal}
  {Physical Review B}\ }\textbf {\bibinfo {volume} {100}},\ \bibinfo {pages}
  {085106} (\bibinfo {year} {2019}{\natexlab{b}})}\BibitemShut {NoStop}%
\bibitem [{\citenamefont {Fischer}\ and\ \citenamefont
  {Berg}(2016)}]{Fischer2016}%
  \BibitemOpen
  \bibfield  {author} {\bibinfo {author} {\bibfnamefont {M.~H.}\ \bibnamefont
  {Fischer}}\ and\ \bibinfo {author} {\bibfnamefont {E.}~\bibnamefont {Berg}},\
  }\bibfield  {title} {\emph {\bibinfo {title} {{Fluctuation and strain effects
  in a chiral $p$-wave superconductor}}},\ }\href {\doibase
  10.1103/PhysRevB.93.054501} {\bibfield  {journal} {\bibinfo  {journal}
  {Physical Review B}\ }\textbf {\bibinfo {volume} {93}},\ \bibinfo {pages}
  {054501} (\bibinfo {year} {2016})}\BibitemShut {NoStop}%
\bibitem [{\citenamefont {Hecker}(2020)}]{Hecker2020-thesis}%
  \BibitemOpen
  \bibfield  {author} {\bibinfo {author} {\bibfnamefont {M.}~\bibnamefont
  {Hecker}},\ }\bibfield  {title} {\emph {\bibinfo {title} {{Fluctuations and
  Nematicity in Unconventional and Topological Superconductors}}},\ }\href@noop
  {} {\bibfield  {journal} {\bibinfo  {journal} {{doctoral thesis \!\!}}\ ,\
  \bibinfo {pages} {Karlsruhe Institute of Technology}} (\bibinfo {year}
  {2020})}\BibitemShut {NoStop}%
\bibitem [{\citenamefont {Xia}\ \emph {et~al.}(2006)\citenamefont {Xia},
  \citenamefont {Maeno}, \citenamefont {Beyersdorf}, \citenamefont {Fejer},\
  and\ \citenamefont {Kapitulnik}}]{Xia2006}%
  \BibitemOpen
  \bibfield  {author} {\bibinfo {author} {\bibfnamefont {J.}~\bibnamefont
  {Xia}}, \bibinfo {author} {\bibfnamefont {Y.}~\bibnamefont {Maeno}}, \bibinfo
  {author} {\bibfnamefont {P.~T.}\ \bibnamefont {Beyersdorf}}, \bibinfo
  {author} {\bibfnamefont {M.~M.}\ \bibnamefont {Fejer}}, \ and\ \bibinfo
  {author} {\bibfnamefont {A.}~\bibnamefont {Kapitulnik}},\ }\bibfield  {title}
  {\emph {\bibinfo {title} {{High Resolution Polar Kerr Effect Measurements of
  ${\mathrm{Sr}}_{2}{\mathrm{RuO}}_{4}$: Evidence for Broken Time-Reversal
  Symmetry in the Superconducting State}}},\ }\href {\doibase
  10.1103/PhysRevLett.97.167002} {\bibfield  {journal} {\bibinfo  {journal}
  {Physical Review Letters}\ }\textbf {\bibinfo {volume} {97}},\ \bibinfo
  {pages} {167002} (\bibinfo {year} {2006})}\BibitemShut {NoStop}%
\end{thebibliography}


\vfill
\pagebreak

\begin{minipage}{\textwidth}
\centering
\LARGE
\textbf{Supplementary Information}
\vspace{2em}
\end{minipage}

\twocolumngrid
In order to evaluate the phase diagram for the above problem, we use a specific model that can, to certain extent, be treated analytically. A valid path to describe three-dimensional anisotropic systems qualitatively consists in formulating the problem in an isotropic model in $2 \!<\! d \!<\! 3$ dimensions \cite{Fernandes2012a}. For simplicity we choose $d = 3 - \varepsilon$ with $\varepsilon = 1/2$. This implies that the momentum integrals take the explicit form $\int_{\vec{q}} \!=\! \Omega_{0} \!\int_{0}^{\infty} q^{3/2}dq$ and $\Omega_{0} \!=\! 5 \pi^{5/4} / [2 \Gamma(9/4)]$ the area of a $(5/2)$-dimensional unit sphere. Furthermore we parametrize the renormalized (bare) transition temperatures of the two primary phases as $T_{c1} \!=\! T_{c0} \!+\! x$, $T_{c2} \!=\! T_{c0} \!-\! x$, with $x$ a tuning parameter such as doping or pressure. The set of equations (6)-(8) in the main text then reads
\begin{align}\label{eq:compact-eq1}
   r_{1} &= (T - T_{c1}) + u \langle \eta^{*}_{1}\rangle \langle \eta^{\phantom{*}}_{1}\rangle + u \mathrm{K}_{1}(r_{1},r_{2},\mu),\\
   \label{eq:compact-eq2}
   r_{2} &= (T - T_{c2}) + u \langle \eta^{*}_{2}\rangle \langle \eta^{\phantom{*}}_{2}\rangle + u \mathrm{K}_{2}(r_{1},r_{2},\mu),\\
   \label{eq:compact-eqmu}
   2\mu &= -i(\langle\eta^{*}_{1}\rangle \langle\eta^{\phantom{*}}_{2}\rangle \!-\! \langle\eta^{\phantom{*}}_{1}\rangle \langle\eta^{*}_{2}\rangle) + 2g \mu \mathrm{K}_{\mu}(r_{1},r_{2},\mu),
\end{align}
with
\begin{align}\label{}
\!\!
   \mathrm{K}_{1}(r_{1},r_{2},\mu) &\equiv\! \int_{\vec{q}} \frac{ 2(r_{2} + \vec{q}^{2}) }{ (r_{1} +  \vec{q}^{2}) (r_{2} +  \vec{q}^{2}) \!-\! (g \mu)^{2}} \!-\! \frac{ 2}{ \vec{q}^{2}} \\
\!\!
   \mathrm{K}_{2}(r_{1},r_{2},\mu) &\equiv\! \int_{\vec{q}} \frac{ 2(r_{1} + \vec{q}^{2}) }{ (r_{1} +  \vec{q}^{2}) (r_{2} +  \vec{q}^{2}) \!-\! (g \mu)^{2}} \!-\! \frac{ 2}{ \vec{q}^{2}} \\
\!\!
   \mathrm{K}_{\mu}(r_{1},r_{2},\mu) &\equiv\! \int_{\vec{q}} \frac{ 2 }{ (r_{1} +  \vec{q}^{2}) (r_{2} +  \vec{q}^{2}) \!-\! (g \mu)^{2}}.
\end{align}
The bare parameters $r_{0,j}$ have been rescaled to $T - T_{cj} \!=\! r_{0,j} + u \int_{\vec{q}} \vec{q}^{-2}$ such that in the absence of phase interactions the primary phases appear at $T_{cj}$, see Eqs. \eqref{eq:compact-eq1} and \eqref{eq:compact-eq2} with $\mu = 0$. An explicit evaluation of the integrals yields
\begin{widetext}
\begin{align}\label{}
   \mathrm{K}_{j}(r_{1},r_{2},\mu) &= \frac{\Omega_{0} \pi}{2\cdot 2^{1/4}} \Bigg\{
   \frac{
      \big[r_{1} \!+\! r_{2} + \sqrt{(r_{1} \!-\! r_{2})^{2} + 4 g^{2} \mu^{2}}\,\big]^{3/4}
      \big[r_{j}^{2} - r_{1}r_{2} + 2 g^{2}\mu^{2} - r_{j} \sqrt{(r_{1} \!-\! r_{2})^{2} + 4 g^{2} \mu^{2}}\big]
      }{
      \big(r_{1} r_{2} - g^{2}\mu^{2}\big)^{3/4} \sqrt{(r_{1}-r_{2})^{2} + 4 g^{2} \mu^{2}}}
      \\ \nonumber & \hspace{5em}
      -
   \frac{
      \big[r_{1} \!+\! r_{2} - \sqrt{(r_{1} \!-\! r_{2})^{2} + 4 g^{2} \mu^{2}}\,\big]^{3/ 4}
      \big[r_{j}^{2} - r_{1}r_{2} + 2 g^{2}\mu^{2} + r_{j} \sqrt{(r_{1} \!-\! r_{2})^{2} + 4 g^{2} \mu^{2}}\big]
      }{
      \big(r_{1} r_{2} - g^{2}\mu^{2}\big)^{3/4} \sqrt{(r_{1}-r_{2})^{2} + 4 g^{2} \mu^{2}}}
   \Bigg\}
   \\
   \mathrm{K}_{\mu}(r_{1},r_{2},\mu) &= \Omega_{0} \pi 2^{1/4}
   \frac{
      \big[r_{1} \!+\! r_{2} + \sqrt{(r_{1} \!-\! r_{2})^{2} + 4 g^{2} \mu^{2}}\,\big]^{1/4}
      - \big[r_{1} \!+\! r_{2} - \sqrt{(r_{1} \!-\! r_{2})^{2} + 4 g^{2} \mu^{2}}\,\big]^{1/4}
      }{
      \sqrt{(r_{1}-r_{2})^{2} + 4 g^{2} \mu^{2}}}
\end{align}
\end{widetext}
The function $\mathrm{K}_{\mu}(r_{1},r_{2})$ in the main text is related to its three-argument cousin via $\mathrm{K}_{\mu}(r_{1},r_{2}) \equiv \mathrm{K}_{\mu}(r_{1},r_{2},0)$. In absence of the composite order, i.e.\ $\mu \!=\! 0$, the functions assume the simple form
\begin{align}\label{}
   \mathrm{K}_{j}(r_{1},r_{2},0) &= - \Omega_{0} 2\pi (r_{j} / 4)^{1/4}
   \\
   \mathrm{K}_{\mu}(r_{1},r_{2},0) &= \Omega_{0} 2\pi \frac{(r_{1} / 4)^{1/4} - (r_{2} / 4)^{1/4}}{r1-r2}
\end{align}
To determine for which doping range the bound-state order appears before the onset temperatures $T_{c1}$ and $T_{c2}$ we evaluate the singularity in $\chi_{\mu}$, i.e.\ for $1 = g \mathrm{K}_{\mu}(r_{1},r_{2},0)$. We find that for $|x| \!<\! x^{*}$ the bound-state order can be favorable, where $x^{*} \!>\! 0$ is the critical doping for which the onset of the composite order coincides with $T_{c1}$ (for $x \!=\! -x^{*}$ it coincides with $T_{c2}$). Solving the coupled set of equations \eqref{eq:compact-eq1}-\eqref{eq:compact-eqmu}, with $x = x^{*}$ and $T \!=\! 1+x^{*}$, yields
\begin{align}\label{eq:r2crit}
   r_{2} &= (4\pi^{4} g^{4} \Omega_{0}^{4} )^{1/3}\\
   x^{*} &= \frac{1}{2}(4\pi^{4} \Omega_{0}^{4} )^{1/3} g^{1/3} (g + u).
\end{align}
Alternatively (scenario 3 in the main text) the system can undergo a first-order transition from $\mu = 0$ to $\mu = (r_{1} r_{2})^{1/2}/g$ into a fully ordered state. This occurs when for the first time a solution to
\begin{align}
   r_{1} &= (T - T_{c1}) + u \mathrm{K}_{1}[r_{1},r_{2},(r_{1} r_{2})^{1/2}/g],\\
   r_{2} &= (T - T_{c2}) + u \mathrm{K}_{2}[r_{1},r_{2},(r_{1} r_{2})^{1/2}/g],\\
   \mu &= g \mu \mathrm{K}_{\mu}[r_{1},r_{2},(r_{1} r_{2})^{1/2}/g],
\end{align}
Some reordering reveals that the bound for a first order transition is independent of $x$ and given by 
\begin{align}\label{}
   T_{c}^{\mathrm{first}}(x) = T^{*}.
\end{align}
We conclude, that all the phase transition lines merge in $(T^{*}, x^{*})$. In the range $|x| < x^{*}$, the appearance of the bound-state order is guaranteed if the susceptibility $\chi_{\mu}$ diverges at temperatures $T_{\mu} > T^{*}$. We have evaluated $T_{\mu}(x)$ for $x = 0$ and in the vicinity of $x^{*}$. We find
\begin{align}\label{eq:Tmu}
   T_{\mu}(0) &= T_{c0} + \frac{1}{4}(\pi^{4} \Omega_{0}^{4})^{1/3} g^{1/3} (g+4u)\\
   \nonumber
   &= T^{*} + \frac{1}{8} (4 \pi^{4} \Omega_{0}^{4})^{1/3}\! g^{1/3} [2^{1/3}(4u \!+\! g) - 4(u \!+\! g)]
\end{align}
and
\begin{align}     
   T_{\mu}(x) &\approx T^{*} - \frac{2g - u}{2(g-u)}(x^{*} \!-\! |x|) \quad \text{for } 1 \!-\! |x|/x^{*} \ll 1.
\end{align}
The slope of $T_{\mu}(x)$ at $x^{*}$ changes sign for $g \!=\! g^{*} \!\equiv\! u/2$, while at this interaction strength the transition $T_{\mu}(0)|_{g=g^{*}}$ at $x \!=\! 0$ is still shadowed by $T_{c}^{\mathrm{first}}$. This tells that, upon reducing $g$, the bound-state phase first becomes favorable near $x^{*}(g)$ and not at the degeneracy point. The bound-state order appears in the entire doping range $|x| < x^{*}$ once $T_{\mu}(0) \!\geq\! T_{c}^{\mathrm{first}}= T^{*}$. Solving $T_{\mu}(0) = T^{*}$ from Eq.\ \eqref{eq:Tmu}, this condition is first satisfied for $g \!=\! g_{c} \!\equiv\! 4 u (2^{1/3} - 1)/(4 - 2^{1/3}) \!\approx\! 0.38 u$. In the intermediate parameter range $g \!\in\! [g_{c}, g^{*}]$ the composite phase exists in two split lobes [defined by the condition $T_{\mu}(x) > T^{*}$], one on each side of the degeneracy line.

The bound-state order occupies the largest temperature range at the degeneracy point, $\delta T = T_{\mu}(0) - T^{*}$, and for the parameter $g_{m} = [(2^{1/3} \!-\! 1) / (4 \!-\! 2^{1/3})] u$, as obtained from maximizing $\delta T$ using Eq.\ \eqref{eq:Tmu}.

For $x \!>\! x^{*}$ the primary order $\langle \eta_{1} \rangle$ appears below $T_{c1} \!=\! T_{c0} + x$ and the second order follows when the susceptibility $\chi_{\mu}$ diverges. While the latter assumes the from in Eq.\ (15) [main text], for most practical purposes, it is well approximated by
\begin{align}
   \chi_{\mu} \approx \mathrm{K}_{\mu}(0,r_{2},0) [1 - g \mathrm{K}_{\mu}(0,r_{2},0)]^{-1},
\end{align}
i.e.\ by neglecting the small correction term $\propto \langle \eta_{1}\rangle^{2} / r_{2}$. The condition for the second phase transition then translates to $T_{c2,\mu}(x) \!\approx\! T_{c2}(x) + 2 x^{*} = T^{*} - (x - x^{*})$, near $x^{*}$.

\vfill

\end{document}